\documentclass[11pt]{article}
\usepackage{amsmath,amsfonts,amssymb,graphicx,epsfig,pdflscape}
\usepackage[usenames]{color}
\usepackage{pstricks}
\numberwithin{equation}{section}

\newcommand{\nc}{\newcommand}
\nc\disp{\displaystyle}
\nc{\fh}{\hat{f}}
\nc{\muh}{\hat{\mu}}
\nc{\nuh}{\hat{\nu}}
\nc{\spos}[2]{\makebox(0,0)[#1]{$\sm{#2}$}}
\nc{\sm}[1]{{\scriptstyle #1}}
\nc{\bib}{\bibitem}
\nc{\al}{\alpha}
\nc{\g}{\gamma}
\nc{\G}{\Gamma}
\nc{\D}{\Delta}
\nc{\eps}{\epsilon}
\nc{\la}{\lambda}
\nc{\La}{\Lambda}
\nc{\var}{\varphi}
\nc{\pa}{\partial}
\nc{\nn}{\nonumber \\ }
\nc{\hf}{\frac{1}{2}}
\nc{\dz}{\frac{dz}{2\pi i}}
\nc{\bin}[2]{\left(\!\!\!\begin{array}{c} {#1}\\ {#2} \end{array}\!\!\!\right)}
\nc{\be}{\begin{equation}}
\nc{\ee}{\end{equation}}
\nc{\bea}{\begin{eqnarray}}
\nc{\eea}{\end{eqnarray}}
\nc{\bra}[1]{\langle {#1}|}
\nc{\ket}[1]{|{#1}\rangle}
\nc{\ketw}[1]{({#1})^{\phantom{a}}_{{\cal W}}}
\nc{\chit}{\raisebox{0.25ex}{$\chi$}}
\nc{\chih}{\raisebox{0.25ex}{$\hat\chi$}}
\nc{\Dti}{\tilde{\D}}
\nc{\Db}{\mbox{\boldmath $D$}}
\nc{\Hb}{\mbox{\boldmath $H$}}
\nc{\Ib}{\mbox{\boldmath $I$}}
\nc{\qb}{\bar{q}}
\nc{\Ac}{\mathcal{A}}
\nc{\Bc}{\mathcal{B}}
\nc{\Cc}{\mathcal{C}}
\nc{\Dc}{\mathcal{D}}
\nc{\Ec}{\mathcal{E}}
\nc{\Fc}{\mathcal{F}}
\nc{\Hc}{\mathcal{H}}
\nc{\Ic}{\mathcal{I}}
\nc{\Jc}{\mathcal{J}}
\nc{\Oc}{\mathcal{O}}
\nc{\Rc}{\mathcal{R}}
\nc{\Vc}{\mathcal{V}}
\nc{\Wc}{\mathcal{W}}
\nc{\Xc}{\mathcal{X}}
\nc{\Yc}{\mathcal{Y}}
\nc{\Zc}{\mathcal{Z}}
\nc{\fus}{\mbox{}\,\hat\otimes\,\mbox{}}
\def\vvdots{\mathinner{\mkern1mu\raise1pt\vbox{\kern7pt\hbox{.}}\mkern2mu
  \raise4pt\hbox{.}\mkern2mu\raise7pt\hbox{.}\mkern1mu}}
\nc{\gauss}[2]{\left[\!\!\begin{array}{c} {#1}\\ {#2} \end{array}\!\!\right]}
\nc{\sbin}[2]{\left\{\!\!\!\begin{array}{c} {#1}\\ {#2} 
\end{array}\!\!\!\right\}}
\nc{\sbinlr}[2]{\Big\langle\!\!\begin{array}{c} {#1}\\ {#2} 
\end{array}\!\!\Big\rangle}
\nc{\bino}[2]{\left(\!\!\begin{array}{c} {#1}\\ {#2} \end{array}\!\!\right)}
\definecolor{lightblue}{rgb}{.61,.61,1}
\definecolor{midblue}{rgb}{.7,.7,1}
\definecolor{lightlightblue}{rgb}{.85,.85,1}
\definecolor{lightestblue}{rgb}{.96,.96,1}
\definecolor{lightpurple}{rgb}{1,.65,1}
\nc{\ch}{{\rm ch}}
\nc{\R}{{\cal R}}
\nc{\dkk}{\delta_{j,\{k,k'\}}^{(2)}}
\nc{\drr}{\delta_{j,\{r,r'\}}^{(2)}}
\nc{\ddkk}{\delta_{j,\{k,k'\}}^{(4)}}
\nc{\dddkk}{\delta_{j,\{k,k'\}}^{(8)}}
\nc{\dnn}{\delta_{j,\{n,n'\}}^{(2)}}
\nc{\ddnn}{\delta_{j,\{n,n'\}}^{(4)}}
\nc{\dddnn}{\delta_{j,\{n,n'\}}^{(8)}}

\definecolor{pink}{rgb}{1,.65,.65}

\begin{document}

\topmargin -5mm
\oddsidemargin 5mm

\setcounter{page}{1}

\vspace{8mm}
\begin{center}
{\huge {\bf Classification of Kac representations}}
\\[.3cm]
{\huge {\bf in the logarithmic minimal models ${\cal LM}(1,p)$}}

\vspace{10mm}
{\LARGE J{\o}rgen Rasmussen}
\\[.3cm]
{\em Department of Mathematics and Statistics, University of Melbourne}\\
{\em Parkville, Victoria 3010, Australia}
\\[.4cm]
{\tt j.rasmussen\,@\,ms.unimelb.edu.au}
\end{center}

\vspace{8mm}
\centerline{{\bf{Abstract}}}
\vskip.4cm
\noindent
For each pair of positive integers $r,s$, 
there is a so-called Kac representation 
$(r,s)$ associated with a Yang-Baxter integrable boundary condition in the lattice
approach to the logarithmic minimal model ${\cal LM}(1,p)$. 
We propose a classification of these representations as finitely-generated submodules of
Feigin-Fuchs modules, and present a conjecture
for their fusion algebra which we call the Kac fusion algebra. 
The proposals are tested using a combination of
the lattice approach and applications of the Nahm-Gaberdiel-Kausch algorithm.
We also discuss how the fusion algebra may be extended by inclusion
of the modules contragredient to the Kac representations, and determine polynomial
fusion rings isomorphic to the conjectured Kac fusion algebra and its contragredient extension.
\renewcommand{\thefootnote}{\arabic{footnote}}
\setcounter{footnote}{0}

\newpage
\tableofcontents

\section{Introduction}

The logarithmic minimal models ${\cal LM}(p,p')$ were introduced in \cite{PRZ0607},
and the present paper concerns these models in the Virasoro picture without
consideration for eventual extensions with respect to some $\Wc$-algebras.
As logarithmic conformal field theories \cite{RS9203,Gur9303,Flohr0111,Gab0111}, 
the models arise in the continuum scaling limit of an infinite family of Yang-Baxter integrable 
lattice models labelled by the pair of coprime integers $p,p'$.
For each pair of positive integers $r,s\in\mathbb{N}$, there is a so-called
Kac representation associated with an integrable boundary condition
in the lattice model \cite{RP0706,RP0707}.
Despite their importance, these Kac representations are in general rather poorly
understood as modules over the Virasoro algebra. Their characters are known empirically
from the lattice approach, but this is in general not sufficient to determine the 
underlying representations.

Fusion can be implemented on the lattice without detailed knowledge of the structure of the
Kac representations. This gives significant insight into the fusion algebra generated from
repeated fusion of the Kac representations and has led to a concrete conjecture
for the so-called fundamental fusion algebra and its representation content 
\cite{RP0706,RP0707}. 
This fundamental fusion algebra is generated from repeated fusion of the
fundamental Kac representations $(2,1)$ and $(1,2)$, but does not involve all
Kac representations.
The lattice implementation of fusion also provides further information
on the structure of the representations themselves, but certain crucial 
questions are left unanswered.
However, the fusion rules should be compatible with the outcome of
the Nahm-Gaberdiel-Kausch (NGK) algorithm \cite{Nahm9402,GK9604},
thus providing an additional tool to determine the fusion rules in ${\cal LM}(p,p')$.
The NGK algorithm has  
played a prominent role in the study of fusion in the so-called augmented $c_{p,p'}$ models
\cite{GK9604,EF0604} as well as in \cite{MR0708,MR0711} on critical percolation 
and related models. Alternative approaches to the computation of fusion rules
in these models are discussed in \cite{Flohr9605,RS0701}.

The logarithmic minimal models are non-rational conformal field theories
as they contain infinitely many Virasoro representations. 
Some of these can be organized in finitely many extended representations
associated with new integrable boundary conditions \cite{PRR0803,RP0804,Ras0805}.
This is referred to as the ${\cal W}$-extended picture of the logarithmic
minimal models, and the extension is believed to be with respect to the triplet 
$\Wc_{p,p'}$ algebra 
\cite{Kausch9510,GK9606,FGST0606}.
Due to their `rational nature', these ${\cal W}$-extended models have been studied
extensively, see \cite{PR1010} and references therein.

Here we consider the infinite family of logarithmic minimal models ${\cal LM}(1,p)$.
We propose a classification of the Kac representations $(r,s)$
for all $r,s\in\mathbb{N}$ as finitely-generated submodules of Feigin-Fuchs modules \cite{FF89}, 
and present a conjecture for their fusion algebra. We thus find that the
only higher-rank representations generated by repeated fusion of the Kac
representations are the rank-2 modules $\R_r^b$ already present in
the fundamental fusion algebra.
The proposals are tested using a combination of
the lattice approach and applications of the NGK algorithm.
Under some natural assumptions about the continuum scaling limit
of the lattice model, some results are in fact {\em exact} rather than conjectural.
We also discuss how the fusion algebra may be extended by inclusion
of the modules contragredient to the Kac representations, and determine polynomial
fusion rings isomorphic to the conjectured Kac fusion algebra and its contragredient extension.

\section{Logarithmic minimal model ${\cal LM}(1,p)$}
\label{SecLM1p}

The logarithmic minimal model ${\cal LM}(1,p)$ is a logarithmic conformal field theory
with central charge
\be
 c=1-6\frac{(p-1)^2}{p}
\label{c}
\ee
Here we are mainly interested in the Virasoro representations 
associated with the boundary conditions appearing in the lattice approach to  
${\cal LM}(1,p)$ as described in \cite{PRZ0607,RP0707}, but consider also other representations.

\subsection{Highest-weight modules}

Before describing the representations associated with boundary conditions, let us
recall some basic facts about highest-weight modules over the Virasoro algebra
with central charge given by (\ref{c}).
For each pair of positive Kac labels $r,s\in\mathbb{N}$, the highest-weight Verma module of 
conformal weight $\D_{r,s}$ is denoted by $V_{r,s}$ where $\D_{r,s}$ is given by the Kac formula
\be
 \D_{r,s}=\frac{(rp-s)^2-(p-1)^2}{4p},\qquad r,s\in\mathbb{Z}
\label{Drs}
\ee
As indicated, it is convenient to consider also negative or 
vanishing Kac labels, in particular when applying the Kac-table symmetries
\be
 \D_{r,s}=\D_{-r,-s},\qquad \D_{r,s}=\D_{r+k,s+kp},\qquad k\in\mathbb{Z}
\ee
The distinct conformal weights appearing in (\ref{Drs}) appear exactly once in the set
\be
 \{\D_{r,s};\,r\in\mathbb{N},\,s\in\mathbb{Z}_{1,p}\}
\ee
where we have introduced the notation
\be
 \mathbb{Z}_{n,m}=\mathbb{Z}\cap[n,m]
\ee

The Verma module $V_{r,s}$ has a proper submodule at Virasoro level $rs$ given by $V_{r,-s}$
(where $V_{r,-s}=V_{r',s'}$ for some $r',s'\in\mathbb{N}$),
allowing us to define the quotient module 
\be
 Q_{r,s}=V_{r,s}/V_{r,-s}
\label{Qrs}
\ee
Its character is given by
\be
 \chit[Q_{r,s}](q)=\frac{q^{\frac{1-c}{24}+\D_{r,s}}}{\eta(q)}\big(1-q^{rs}\big)
  =\frac{q^{(rp-s)^2/4p}}{\eta(q)}\big(1-q^{rs}\big)
\label{Qchar}
\ee
where $q$ is the modular nome while the Dedekind eta function is given by
\be
 \eta(q)=q^{1/24}\prod_{m\in\mathbb{N}}(1-q^m)
\ee
This module is in general not irreducible. The
irreducible highest-weight module $M_{r,s}$ of conformal weight $\D_{r,s}$
is obtained by quotienting out the {\em maximal} proper submodule of $V_{r,s}$, and we denote its
character by 
\be
 \ch_{r,s}(q)=\chit[M_{r,s}](q)
\ee
For
\be
 s=s_0+kp,\qquad s_0\in\mathbb{Z}_{1,p-1};\quad k\in\mathbb{N}_0
\label{s}
\ee
the character of the quotient module $Q_{r,s}$ can be written as
\be
 \chit[Q_{r,s}](q)=\sum_{j=0}^{\min(2r-1,2k)}\ch_{r+k-j,(-1)^j s_0+(1-(-1)^j)p/2}(q)
\ee

\subsection{Kac representations}

There is a so-called Kac representation $(r,s)$ for each pair of positive Kac labels
$r,s\in\mathbb{N}$. It is associated with a Yang-Baxter integrable boundary condition in the lattice
approach to ${\cal LM}(1,p)$ \cite{PRZ0607,RP0707} and arises in the continuum scaling limit.
As we will discuss, these Kac representations can be irreducible, fully reducible
or reducible yet indecomposable as modules over the Virasoro algebra. 
They are all of rank 1 as the dilatation generator 
(the Virasoro mode $L_0$) is found to be diagonalizable.
Empirically, the Virasoro character of the Kac representation $(r,s)$ is identical to the character
(\ref{Qchar}) of the quotient module $Q_{r,s}$
\be
 \chit_{r,s}(q)=\chit[Q_{r,s}](q)
\label{char}
\ee
Aside from its character and its rank-1 nature, it is not, however, a priori clear from the lattice 
what type of Virasoro module the Kac representation $(r,s)$ actually is. 
A typical dilemma is the distinction between a reducible yet indecomposable 
module and the direct sum of its irreducible subfactors (subquotients). 
By construction, the indecomposable module has
the same character as the direct sum but they are nevertheless inequivalent due
to the indecomposable nature of the former. The situation can be 
rather intricate, as we will argue below, since some Kac representations are found to be
non-highest-weight representations. Our assertion is that they can all be viewed as
finitely-generated submodules
of Feigin-Fuchs modules, see (\ref{emb}) and (\ref{23013}), for example.
In particular, despite the character identity (\ref{char}),
we thus assert that $(r,s)$ and $Q_{r,s}$ in general differ as representations.

It follows from the character expressions that 
$(r,s)$ is irreducible for $s\in\mathbb{Z}_{1,p}$ and that $(1,kp)$ is irreducible for 
$k\in\mathbb{N}$, thus giving rise to the identifications $(1,rp)\equiv(r,p)$ \cite{RP0707}.
These are the only irreducible Kac representations.
Following Section 4.2 in \cite{RP0707},
one deduces that the Kac representation $(r,kp)$ is fully reducible
\be
 (r,kp)=\bigoplus_{j=|r-k|+1,\,{\rm by}\,2}^{r+k-1}(j,p)
\label{pkpk}
\ee
The remaining Kac representations were not fully characterized in \cite{RP0707}.
Below, we offer a conjecture for the classification of the {\em full} set of Kac representations.

\subsection{Rank-2 representations}

The infinite family 
\be
 \{\R_r^b\,;\, r\in\mathbb{N},\,b\in\mathbb{Z}_{1,p-1}\}
\label{r2}
\ee
of reducible yet indecomposable modules of rank 2 arises
from repeated fusion of {\em irreducible} Kac representations \cite{RP0707}.
This follows by isolating $\R_r^b$ in
\be
 (1,b+1)\otimes(1,rp)=\bigoplus_\beta^b\R_r^\beta,\qquad b\in\mathbb{Z}_{1,p-1}
\ee
for example, as $b$ increases from 1 to $p-1$.
Here we have introduced the summation convention
\be
 \bigoplus_n^N R_n=\bigoplus_{n=\eps(N),\,\mathrm{by}\,2}^N R_n,\qquad 
 \eps(N)=\frac{1-(-1)^N}{2}=N\ (\mathrm{mod}\ 2)
 \label{parity}
\ee
and extended the notation $\R_r^b$ by writing
\be
 \R_r^0\equiv(1,rp)\equiv(r,p)
\label{R0}
\ee
for the irreducible rank-1 module $(r,p)$.
The rank-2 module $\R_r^b$ is characterized by the structure diagram
\psset{unit=.25cm}
\setlength{\unitlength}{.25cm}
\be
\mbox{}
\hspace{-1cm}
 \mbox{
 \begin{picture}(13,6)(0,3.5)
    \unitlength=1cm
  \thinlines
\put(-1.8,1){$\R_1^b:$}
\put(1,2){$M_{2,b}$}
\put(-.4,1){$M_{1,p-b}$}
\put(2,1){$M_{1,p-b}$}
\put(1.05,1){$\longleftarrow$}
\put(1.65,1.5){$\nwarrow$}
\put(0.65,1.5){$\swarrow$}
 \end{picture}
},
\hspace{3cm}
 \mbox{
 \begin{picture}(13,6)(0,3,5)
    \unitlength=1cm
  \thinlines
\put(-1.8,1){$\R_r^b:$}
\put(0.8,2){$M_{r+1,b}$}
\put(-0.4,1){$M_{r,p-b}$}
\put(2,1){$M_{r,p-b}$}
\put(0.8,0){$M_{r-1,b}$}
\put(1.05,1){$\longleftarrow$}
\put(1.65,1.5){$\nwarrow$}
\put(0.65,1.5){$\swarrow$}
\put(1.65,0.5){$\swarrow$}
\put(0.65,0.5){$\nwarrow$}
\put(3.5,1){$,\qquad r\in\mathbb{Z}_{\geq2}$}
 \end{picture}
}
\label{Remb}
\\[0.8cm]
\ee
where an arrow from the irreducible subfactor $M$ to the irreducible subfactor $M'$
indicates that vectors in $M$ are mapped not only to vectors in $M$ itself but also to vectors
in $M'$ by the action of the Virasoro algebra. 
An arrow from one copy of $M$ to another copy of $M$ indicates that
$L_0$ is non-diagonalizable and that the module is of rank 2. Representations of rank $\rho>2$
are not present here, but would otherwise require $\rho$ copies of a given 
irreducible subfactor suitably arranged in a chain and connected by aligned arrows.
The character of the rank-2 module $\R_r^b$ follows from the structure diagram (\ref{Remb})
and is given by
\be
 \chit[\R_r^b](q)
  =(1-\delta_{r,1})\ch_{r-1,b}(q)+2\ch_{r,p-b}(q)+\ch_{r+1,b}(q)
\ee
According to the fusion algebra conjectured in Section~\ref{SecFull}, 
no additional rank-2 modules nor higher-rank modules are
generated from repeated fusion of the {\em full} set of Kac representations $(r,s)$.

\subsection{Reducible yet indecomposable Kac representations}
\label{SecRed}

There is a pair of Feigin-Fuchs modules corresponding to each Verma module $V_{r,s}$.
We denote them by $F^{\to}_{r,s}$ and $F^{\gets}_{r,s}$, and 
they can be constructed by reversing every second arrow in the structure diagram for $V_{r,s}$. 
The arrow on $F^\to_{r,s}$ indicates that vectors in $M_{r,s}$ are mapped not only to vectors
in $M_{r,s}$ itself but also {\em to} vectors in the next
subfactor by the action of the Virasoro algebra. Similarly, the arrow on $F^\gets_{r,s}$ 
indicates that vectors in $M_{r,s}$ can be reached {\em from} vectors in the next subfactor.
Likewise, we can associate a pair of finitely-generated
Feigin-Fuchs modules to every quotient module $Q_{r,s}$.
For $2r-1<2k$, where $s=s_0+kp$,
the Feigin-Fuchs modules corresponding to $Q_{r,s}$ are characterized by the structure diagrams
\be
\begin{array}{rcl}
 &Q^{\to}_{r,s}:&\quad
  M_{k-r+1,p-s_0}\to M_{k-r+2,s_0}\gets M_{k-r+3,p-s_0}\to\ldots\gets M_{k+r-1,p-s_0}\to 
    M_{k+r,s_0}
 \\[.5cm]
 &Q^{\gets}_{r,s}:&\quad
  M_{k-r+1,p-s_0}\gets M_{k-r+2,s_0}\to M_{k-r+3,p-s_0}\gets\ldots\to M_{k+r-1,p-s_0}\gets 
    M_{k+r,s_0}
\end{array}
\ee
For $2r-1>2k$, the Feigin-Fuchs modules corresponding to $Q_{r,s}$ are characterized by the
structure diagrams
\be
\begin{array}{rcl}
 &Q^{\to}_{r,s}:&\quad
  M_{r-k,s_0}\to M_{r-k+1,p-s_0}\gets M_{r-k+2,s_0}\to\ldots\to M_{r+k-1,p-s_0}\gets M_{r+k,s_0}
 \\[.5cm]
 &Q^{\gets}_{r,s}:&\quad
  M_{r-k,s_0}\gets M_{r-k+1,p-s_0}\to M_{r-k+2,s_0}\gets\ldots\gets M_{r+k-1,p-s_0}\to M_{r+k,s_0}
\end{array}
\ee
By construction, we have
\be
 \chit[Q^{\to}_{r,s}](q)= \chit[Q^{\gets}_{r,s}](q)= \chit[Q_{r,s}](q)
\ee
In all cases, the Feigin-Fuchs modules $Q^{\to}_{r,s}$ and $Q^{\gets}_{r,s}$ 
are {\em contragredient} to each other
where the contragredient module $A^\ast$ to a module $A$ is obtained by reversing all 
structure arrows between
its irreducible subfactors. It follows that $\chit[A^\ast](q)=\chit[A](q)$ and that $A^{\ast\ast}=A$.

Since we know the structure of the Kac representation $(r,s)$ for $s\leq p$ (irreducible) or 
$s=kp$ (fully reducible), we now consider the cases where $s$ is of the form (\ref{s}) for $k\geq1$.
\\[.2cm]
{\bf Highest-weight assumption.}\quad 
The Kac representation $(1,s_0+kp)$ is the indecomposable highest-weight module
$Q^\to_{1,s_0+kp}$, that is,
\be
 (1,s_0+kp)=Q^\to_{1,s_0+kp}=Q_{1,s_0+kp}=V_{1,s_0+kp}/V_{1,s_0+(k+2)p}:\qquad 
    M_{1,s_0+kp}\to M_{1,(k+2)p-s_0}
\label{1s}
\ee
It is emphasized that, a priori, this Kac representation could be the direct sum 
of its two irreducible subfactors or contragredient to the highest-weight module (\ref{1s}).
Consistency of the fusion algebra excludes the first possibility, though. 
To appreciate this, let us initially compare certain fusion properties of the Kac
representation $(1,p+1)$ with the similar properties of the direct sum $(1,p-1)\oplus(2,1)$ of 
its constituent subfactors. According to the fundamental fusion algebra \cite{RP0707}, we have
\be
 \big[(1,p-1)\oplus(2,1)\big]\otimes\big[(1,p-1)\oplus(2,1)\big]
  =2(1,1)\oplus(3,1)\oplus2(2,p-1)\oplus\bigoplus_\beta^{p-3}\R_1^\beta
\label{1p21}
\ee
On the other hand, it is observed from the lattice that the decomposition of the
fusion product $(1,p+1)\otimes(1,p+1)$ for small $p$ contains rank-2 Jordan cells
linking the two copies of the irreducible subfactor $M_{1,1}=(1,1)$. This is incompatible
with (\ref{1p21}) as the former indicates the presence of the rank-2 module $\R_1^{p-1}$
(\ref{Remb}) in the decomposition, in accordance with the conjectured fusion rule (\ref{fullver2})
\be
 (1,p+1)\otimes(1,p+1)=(1,2p+1)\oplus\bigoplus_\beta^{p-1}\R_1^\beta
\ee
More generally, the lattice approach 
provides similar evidence for the indecomposability of $(1,s_0+kp)$. We thus find that
\be
 (1,s_0+kp)\neq(k,p-s_0)\oplus(k+1,s_0)
\ee
since the lattice approach indicates the presence of $\R_1^{p-1}$ in the decomposition
of the fusion product $(1,s_0+kp)\otimes(1,s_0+kp)$, while the decomposition of the
fusion product of $(k,p-s_0)\oplus(k+1,s_0)$ with itself follows from the fundamental 
fusion algebra 
\be
 \big[(k,p-s_0)\oplus(k+1,s_0)\big]\otimes\big[(k,p-s_0)\oplus(k+1,s_0)\big]=2(1,1)\oplus\ldots
\ee
and contains two {\em unlinked} copies of $M_{1,1}$.

We are still faced with the problem of identifying the Kac representations 
$(1,s_0+kp)$ as highest-weight modules or as the corresponding contragredient modules.
As indicated, here we {\em assume} that they are highest-weight modules and then study the
implications of this assumption. We will nevertheless return to this question in 
Section~\ref{SecContra} and Section~\ref{SecFusion}. 
\\[.2cm]
\noindent
{\bf Structure conjecture.}\quad For $s=s_0+kp$, $k\in\mathbb{N}$,
the Kac representation $(r,s)$ is the Feigin-Fuchs module
\be
 (r,s)=\begin{cases} Q^{\to}_{r,s},\ &2r-1<2k
  \\[.2cm]
  Q^{\gets}_{r,s},\ &2r-1>2k
 \end{cases}
\label{emb}
\ee
Below, we present arguments in support of this conjecture by combining
results from the lattice approach with results from applications of the NGK algorithm based 
on (\ref{1s}).

The range for $s_0$ in (\ref{emb}) can be extended from $\mathbb{Z}_{1,p-1}$ to 
$\mathbb{Z}_{0,p-1}$ such that $s$ can be any positive integer
$s\in\mathbb{N}$ (where we exclude $s_0=k=0$ for which $s=0$).
For $s_0=0$, the structure diagrams associated with $Q^\to_{r,s}$ and $Q^\gets_{r,s}$
in (\ref{emb}) are separable (degenerate) and the modules are fully reducible
\be
 Q_{r,kp}^\to=Q_{r,kp}^\gets=Q_{r,kp}=\bigoplus_{j=|r-k|+1,\,\mathrm{by}\,2}^{r+k-1}M_{j,p}
\label{s00}
\ee
in accordance with (\ref{pkpk}).
It is noted that this decomposition is symmetric in $r$ and $k$.

In retrospect, we could have {\em defined} a Kac representation $(r,s)$ mathematically,
for general $r,s\in\mathbb{N}$, as the finitely-generated Feigin-Fuchs submodule (\ref{emb}) where
\be
 s=s_0+kp,\qquad s_0\in\mathbb{Z}_{0,p-1},\quad k\in\mathbb{N}_0
\ee
{}From the lattice, we would then conjecture that the Virasoro modules
associated with the aforementioned boundary conditions
are Kac representations in the mathematical sense just given. 
A major goal of the present work is indeed to collect evidence for this conjecture.
It is recalled that we are working under the assumption that the modules $(1,s_0+kp)$
are highest-weight modules.

\subsubsection{Evidence for the structure conjecture}

Fusion can be implemented on the lattice
without detailed knowledge of the structure of the Kac representations.
The Kac representation $(r,s)$ itself is actually constructed by fusing the `horizontal' 
Kac representation $(r,1)$ with the `vertical' Kac representation $(1,s)$
\be
 (r,s)=(r,1)\otimes (1,s)
\label{r11s}
\ee
Under the assumption (\ref{1s}), we have applied the NGK algorithm to many fusion 
products of this kind and they all corroborate the structure conjecture (\ref{emb}).
Some of our findings and observations are summarized in the following with additional
details deferred to Appendix~\ref{AppNGK}.

In the decomposition of a fusion product examined using the NGK algorithm,
the vectors appearing at Nahm level 0 are the ones which are not the image of
negative Virasoro modes. These 
vectors\footnote{These vectors actually span a subspace, but it is convenient 
to think in terms of a set of basis vectors.}
constitute the minimal set of vectors from which the entire 
(decomposable or indecomposable) module, arising as the result of the fusion product,
can be generated by the action of negative Virasoro modes only.
It thus suffices to analyze a fusion product at Nahm level 0 in order to
identify this minimal set of vectors. This knowledge is then sufficient to
distinguish between a highest-weight module like (\ref{1s}) and its contragredient
module. Indeed, the minimal set of vectors associated with the highest-weight
module in (\ref{1s}) consists of only one vector, namely $\ket{\D_{1,s_0+kp}}$,
whereas the minimal set associated with the contragredient module
consists of the two vectors $\ket{\D_{1,s_0+kp}}$ and $\ket{\D_{1,(k+2)p-s_0}}$.

Once we know this minimal set, we can use our knowledge of the character $\chit_{r,s}(q)$
to deduce the number and conformal weights of the vectors appearing at higher Nahm levels. 
This is very helpful when determining the otherwise evasive spurious subspaces 
appearing in the NGK algorithm, see Appendix~\ref{AppNGK}.
\\[.2cm]
\noindent
{\bf Singular vector conjecture.}\quad
With the normalization convention for singular vectors used in Appendix~\ref{AppSing},
we conjecture that at Nahm level 0 in the fusion product $(2,1)\otimes(1,s)$
\be
 \ket{\D_{2,1}}\times\ket{\la_{1,s}}=-\big(\prod_{j=1}^{s-1}\frac{(p+j)(p-j)}{p}\big)
  \big\{L_{-1}\times I+\tfrac{s-1}{2}I\times I\big\}\ket{\D_{2,1}}\times\ket{\D_{1,s}}
\label{D21}
\ee
We have verified this remarkably simple expression explicitly for $s\leq6$.
The action of the co-multiplication of $L_0$
on the corresponding two-dimensional initial vector space is given by 
\bea
 \D(L_0)\ket{\D_{2,1}}\times\ket{\D_{1,s}}
   &=&(\D_{2,1}+\D_{1,s})\ket{\D_{2,1}}\times\ket{\D_{1,s}}
   +L_{-1}\ket{\D_{2,1}}\times\ket{\D_{1,s}}  \nn
 \!\!\!\!\!\!\D(L_0)L_{-1}\ket{\D_{2,1}}\times\ket{\D_{1,s}}
  &=&p\D_{1,s}\ket{\D_{2,1}}\times\ket{\D_{1,s}}
  +(\D_{2,1}+\D_{1,s}+1-p)L_{-1}\ket{\D_{2,1}}\times\ket{\D_{1,s}}
\label{DL0}
\eea
It follows readily from (\ref{D21}) that a spurious subspace at Nahm level 0
is generated by setting the singular vector $\ket{\la_{1,s}}=0$ if and only if $s\leq p$,
in which case this subspace is one-dimensional.
For $s\leq p$, the matrix realization of
$\D(L_0)$ is therefore one-dimensional and is given by $\D_{2,s}$, reflecting
that the Kac representation $(2,s)$ is irreducible for all $s\leq p$.
For $s>p$, it follows from (\ref{DL0}) that the two-dimensional matrix realization of
$\D(L_0)$ is diagonalizable and has 
eigenvalues $\D_{1,s-p}$ and $\D_{1,s+p}$. 
For $s=kp$, this is in accordance with the decomposition
(\ref{pkpk}), while for $s=s_0+kp$, it is in accordance with the structure conjecture (\ref{emb}).

At Nahm level 0, 
we have confirmed the structure conjecture (\ref{emb}) in many cases. In some of these,
we have continued the analysis to higher Nahm level and always with affirmative results.
Details of the analysis for $(2,3)$ in critical dense polymers ${\cal LM}(1,2)$ appear 
in Appendix~\ref{App23} and are summarized by the structure diagram
\be
 (2,3)=Q_{2,3}^\gets:\qquad \Vc(0)\gets\Vc(1)\to\Vc(3)
\label{23013}
\ee
where $\Vc(\D)$ denotes the irreducible highest-weight module of conformal weight $\D$.

\subsection{Contragredient Kac representations}
\label{SecContra}

We recall our working assumption (\ref{1s}) that the reducible yet indecomposable
Kac representation $(1,s_0+kp)$ is the highest-weight module 
$Q^{\to}_{1,s_0+kp}$ and not its contragredient module $Q^{\gets}_{1,s_0+kp}$.
It then follows from the structure diagram (\ref{Remb}) that the rank-2 module
$\R_r^b$ admits the short exact sequence
\be
 0\to (1,rp-b)\to\R_r^b\to(1,rp+b)\to0
\ee
It also admits the short exact sequence
\be
 0\to(r,p-b)\to\R_r^b\to(r,p+b)\to0
\ee
in terms of the (for $r\neq1$) reducible yet indecomposable non-highest-weight
module $(r,p+b)$.

As Virasoro modules, the Feigin-Fuchs modules contragredient to the ones appearing
in (\ref{emb}), namely
\be
 (r,s)^\ast=\begin{cases} Q^{\gets}_{r,s},\ &2r-1<2k
  \\[.2cm]
  Q^{\to}_{r,s},\ &2r-1>2k
 \end{cases}
\label{Cemb}
\ee
are perfectly well defined. An immediate application is to provide alternative characterizations
of the rank-2 module $\R_r^b$ in terms of short exact sequences as we have
\be
\begin{array}{c}
 0\to(1,rp+b)^\ast\to\R_r^b\to(1,rp-b)^\ast\to0
\\[.3cm] 
 0\to(r,p+b)^\ast\to\R_r^b\to(r,p-b)^\ast\to0
\end{array}
\ee
noting that the rank-2 modules are invariant under reversal of structure arrows
\be
 (\R_r^b)^\ast=\R_r^b
\ee
For $r>1$, the rank-2 module $\R_r^b$ thus admits four independent 
non-trivial short exact sequences.
This is in accordance with the structure diagram (\ref{Remb}) for $\R_r^b$ as it follows
from the diagram that $\R_r^b$ has four inequivalent proper submodules.
We also note that a fully reducible (in particular irreducible) 
module is identical to its contragredient module
\be
 (r,s)^\ast=(r,s),\qquad r\in\mathbb{N};\quad s\in\mathbb{Z}_{1,p-1}\cup p\mathbb{N}
\label{rsast}
\ee

It seems natural to expect that the category or family of representations appearing in
the full-fledged logarithmic conformal field theory ${\cal LM}(1,p)$
is closed under reversal of structure arrows in the sense that the contragredient
module to a module in the category is also in the category. 
Above, we have only considered the Virasoro modules
associated with boundary conditions \cite{PRZ0607,RP0707},
namely the Kac representations $(r,s)$ and the rank-2 modules $\R_r^b$.
As already mentioned, these rank-2 modules are invariant under reversal of structure arrows,
whereas the only invariant Kac representations are the fully reducible ones 
(including the irreducible ones).
As a consequence of the indicated expectation, the 
{\em contragredient Kac representations} (\ref{Cemb}) should also be members of the invariant category.
This situation resembles the logarithmic minimal models ${\cal LM}(p,p')$ in the
so-called $\Wc$-extended picture \cite{PRR0803,Ras0805}
in which the modules associated
with boundary conditions only constitute a subcategory of the
full category if $p>1$. This idea was originally put forward and examined
in \cite{Ras0812} and has since been studied in more detail
\cite{GRW0905,Ras0906,Wood0907,GRW1008}.
In Section~\ref{SecContraFusion}, we discuss how the fusion algebra
generated by the Kac representations may be extended by the inclusion
of the contragredient Kac representations.

\section{Fusion algebras}
\label{SecFusion}

\subsection{Fundamental fusion algebra}

There are infinitely many fusion (sub)algebras associated with ${\cal LM}(1,p)$. 
The {\em fundamental fusion algebra} \cite{RP0707}
\be
 \big\langle (1,1),(2,1),(1,2)\big\rangle
\label{fund}
\ee
in particular, is generated from the two fundamental Kac representations $(2,1)$ and $(1,2)$
in addition to the identity $(1,1)$.
This fusion algebra involves all the irreducible Kac representations
and all the rank-2 representations (\ref{r2}).
On the other hand, no reducible yet indecomposable Kac representations 
arise as the result of repeated fusion of the fundamental Kac representations.
The fundamental fusion algebra has two canonical subalgebras
\be
 \big\langle(1,1),(2,1)\big\rangle,\qquad\big\langle(1,1),(1,2)\big\rangle
\label{horver}
\ee

\subsection{Kac fusion algebra}
\label{SecFull}

The {\em Kac fusion algebra} is generated by the {\em full} set of Kac representations 
\be
 \big\langle(r,s);\ r,s\in\mathbb{N}\big\rangle
\label{full}
\ee
and its description is a main objective of this work.
To appreciate this fusion algebra, it is instructive to examine
its vertical component 
\be
 \big\langle(1,s);\ s\in\mathbb{N}\big\rangle
\label{fullver}
\ee
which is characterized by the fusion rules of the vertical component 
$\langle(1,1),(1,2)\rangle$ of the fundamental fusion algebra
supplemented by the fusion rules involving the reducible yet indecomposable
Kac representations $(1,s_0+kp)$. To describe (\ref{fullver}), we introduce
the sign function
\be
 \mathrm{sg}(n)=\begin{cases} 1,\ &n>0\\ -1,\ &n<0 \end{cases}
\ee
Since this function only appears in conjunction with certain constraints,
the value $\mathrm{sg}(0)$ turns out to be immaterial.
\\[.2cm]
\noindent
{\bf Fusion conjecture.}\quad The vertical component of the Kac fusion algebra
satisfies
\be
 \big\langle(1,s);\ s\in\mathbb{N}\big\rangle
 =\big\langle(1,b+kp),\R_r^b;\ b\in\mathbb{Z}_{0,p-1},\, 
   k\in\mathbb{N}_0,\, r\in\mathbb{N}\big\rangle
\label{fullver0}
\ee
where we recall $\R_r^0\equiv(1,rp)$ and set $(1,0)\equiv\R_0^\beta\equiv0$, 
and is characterized by the fusion rules\footnote{This revises the conjecture in \cite{PR0610} for 
the decomposition of the fusion product $(1,2j_1-1)\otimes(1,2j_2-1)$ in ${\cal LM}(1,2)$,
see also Appendix~\ref{AppCrit}.}
\bea
 (1,b+kp)\otimes(1,b'+k'p)&=&\bigoplus_{j=|k-k'|+1,\,\mathrm{by}\,2}^{k+k'-1}
   \!\!\bigoplus_{\beta}^{p-|b-b'|-1}\R_{j}^\beta
   \oplus\bigoplus_{j=|k-k'+\mathrm{sg}(b-b')|+1,\,\mathrm{by}\,2}^{k+k'}
   \!\!\bigoplus_{\beta}^{|b-b'|-1}\R_{j}^\beta\nn
 &\oplus&\bigoplus_{\beta}^{b+b'-p-1}\R_{k+k'+1}^\beta
   \oplus\bigoplus_{\beta=|b-b'|+1,\,\mathrm{by}\,2}^{p-|p-b-b'|-1}(1,\beta+(k+k')p)\nn
 \R_{r}^{b}\otimes(1,b'+k'p)&=&\bigg(
  \bigoplus_{j=|r-k'|+1,\,\mathrm{by}\,2}^{r+k'-1}
  \!\!\bigoplus_{\beta}^{p-|b-b'|-1}\R_{j}^\beta
  \oplus\bigoplus_{j=|r-k'-1+\mathrm{sg}(p-b-b')|+1,\,\mathrm{by}\,2}^{r+k'-\mathrm{sg}(p-b-b')}
  \!\!\bigoplus_{\beta}^{|p-b-b'|-1}\R_{j}^\beta    \nn
 &\oplus&\bigoplus_{j=|r-k'-1|+1,\,\mathrm{by}\,2}^{r+k'-2}
  \!\!\bigoplus_{\beta}^{p-|p-b-b'|-1}\R_{j}^\beta
  \oplus\bigoplus_{j=|r-k'+\mathrm{sg}(b-b')|+1,\,\mathrm{by}\,2}^{r+k'}
  \!\!\bigoplus_{\beta}^{|b-b'|-1}\R_{j}^\beta\nn
 &\oplus&\bigoplus_{\beta}^{b'-b-1}\R_{r+k'}^\beta
  \oplus\bigoplus_{\beta=|b-b'|+1,\,\mathrm{by}\,2}^{p-|p-b-b'|-1}\R_{r+k'}^\beta
  \bigg)/(1+\delta_{b,0})   \nn
  \R_r^b\otimes\R_{r'}^{b'}&=&\bigg(
   \bigoplus_{j=|r-r'|,\,\mathrm{by}\,2}^{r+r'}
   \big(2-\delta_{j,|r-r'|}\big)
   \Big\{\bigoplus_{\beta}^{|b-b'|-1}\oplus\,\big(1-\delta_{j,r+r'}\big)\!\!\bigoplus_{\beta}^{p-|p-b-b'|-1}
   \Big\}\R_j^\beta\nn
 &\oplus&\Big\{
   \bigoplus_{j=|r-r'-1+\mathrm{sg}(p-b-b')|+1,\,\mathrm{by}\,2}^{r+r'-\mathrm{sg}(p-b-b')}
  \oplus
   \bigoplus_{j=|r-r'+1-\mathrm{sg}(p-b-b')|+1,\,\mathrm{by}\,2}^{r+r'-1}\Big\}
  \bigoplus_{\beta}^{|p-b-b'|-1}\R_{j}^\beta\nn
 &\oplus&
 \bigoplus_{\beta=|b-b'|+1,\,\mathrm{by}\,2}^{p-|p-b-b'|-1}\R_{r+r'}^\beta
  \oplus\bigoplus_{\beta=|p-b-b'|+1,\,\mathrm{by}\,2}^{p-|b-b'|-1}\R_{r+r'-1}^\beta
  \oplus\bigoplus_{\beta}^{p-b-b'-1}\R_{r+r'-1}^\beta
   \nn
  &\oplus&
   \bigoplus_{j=|r-r'|+1,\,\mathrm{by}\,2}^{r+r'-1}\big(2-\delta_{j,r+r'-1}\big)
    \bigoplus_{\beta}^{p-|b-b'|-1}\R_j^\beta
  \bigg)/\{(1+\delta_{b,0})(1+\delta_{b',0})\}
\label{fullver2}
\eea
The divisions by $(1+\delta_{b,0})$ and $(1+\delta_{b',0})$ ensure that the fusion
rules for $\R_r^0$ match those for $(1,rp)$.
Evidence for this fusion conjecture is presented in Section~\ref{SecEviFusionLattice}
and Section~\ref{SecEviFusionNGK}.

Mnemonically, the fusion rules (\ref{fullver2})
are reconstructed straightforwardly using the underlying $sl(2)$ structure \cite{RP0707}.
This structure is evident from the lattice where defects can be
annihilated in pairs thus implying that the fusion product of two Kac representations
$(1,s)$ and $(1,s')$ can be decomposed, up to indecomposable structures, 
as a sum of Kac representations
\be
 (1,s)\otimes(1,s')=(1,|s-s'|+1)\stackrel{\mbox{?}}{\oplus}(1,|s-s'|+3)\stackrel{\mbox{?}}{\oplus}
   \ldots\stackrel{\mbox{?}}{\oplus}(1,s+s'-1)
\label{ss}
\ee 
The question marks indicate that the sums can be {\em direct} or {\em indecomposable}.
The $sl(2)$ structure of the fusion product $(1,s)\otimes (1,s')$ is thus encoded in the
character decomposition
\bea
 \chit\!\left[(1,s)\otimes(1,s')\right]\!(q)
  &=&\chit\!\left[(1,|s-s'|+1)\oplus(1,|s-s'|+3)\oplus\ldots\oplus(1,s+s'-1)\right]\!(q)\nn
  &=&\sum_{s''=|s-s'|+1,\ \!{\rm by}\ \!2}^{s+s'-1}\chit_{1,s''}(q)\nn
  &=&\sum_{t=0}^{\min\{s,s'\}-1}\chit_{1,s+s'-2t-1}(q)
\label{chardec}
\eea
Following the discussion of short exact sequences in Section~\ref{SecContra},
we may view the rank-2 module $\R_r^b$ as an indecomposable combination
of the two Kac representations $(1,rp-b)$ and $(1,rp+b)$, that is, 
\be
 \R_r^b=(1,rp-b)\oplus_i(1,rp+b)
\label{Rrb}
\ee
Utilizing this, we introduce the `forgetful functor' $\Fc$ by
\be
 \Fc[(1,s)]=(1,s),\qquad \Fc[\R_r^b]=(1,rp-b)\oplus(1,rp+b),\qquad
  \Fc[\Ac\otimes\Bc]=\Fc[\Fc[\Ac]\otimes\Fc[\Bc]]
\label{F}
\ee
and apply it to the various fusion products such as
\be
 \Fc[(1,s)\otimes(1,s')]=\bigoplus_{s''=|s-s'|+1,\,\mathrm{by}\,2}^{s+s'-1}(1,s'')
\label{Ffus}
\ee
We note that applying $\Fc$ does not correspond to moving to the Grothendieck ring
associated with characters since we are working here with the reducible yet indecomposable
Kac representations $(1,rp\pm b)$.
Clearly, $\Fc$ does not have an inverse, but on fusion products, we can
devise a prescription that `reintroduces' the rank-2 modules in a unique and well-defined way.
To describe this prescription, let us consider the fusion product $(1,s)\otimes(1,s')$
in (\ref{Ffus}) and initially focus on the Kac representation $(1,s_1'')$ with minimal Kac
label $s_1''=|s-s'|+1$. Depending on $p$, this will appear as the submodule $(1,rp-b)$ of
the rank-2 module $\R_r^b$ if and only if the matching module $(1,rp+b)$ also appears in the
decomposition in (\ref{Ffus}). If not, the Kac representation $(1,s_1'')$ will appear `by itself' in
the decomposition of the fusion product. 
Having completed the examination of $(1,s_1'')$, we remove it
together with its potential partner $(1,rp+b)$ from the direct sum
in (\ref{Ffus}) and repeat the analysis for $(1,s_2'')$ corresponding to the new minimal Kac 
label $s_2''$. This algorithm is continued until all the Kac representations in (\ref{Ffus})
have been accounted for. 
This prescription also works for more complicated fusion products than $(1,s)\otimes(1,s')$ 
and always yields a unique and well-defined result, namely the fusion rules
given in (\ref{fullver2}).
Loosely speaking, the prescription 
corresponds to writing the decomposition of a fusion product in terms of
Kac representations and then forming rank-2 modules whenever possible, starting
with the lowest Kac label and moving up.

\subsubsection{Full Kac fusion algebra}

To describe the {\em full} Kac fusion algebra, not just its vertical component (\ref{fullver0}), 
we note that the horizontal component
$\langle(r,1);\, r\in\mathbb{N}\rangle$ is characterized by the ordinary $sl(2)$ fusion rules
\be
 (r,1)\otimes(r',1)=\bigoplus_{r''=|r-r'|+1,\,\mathrm{by}\,2}^{r+r'-1}(r'',1),\qquad r,r'\in\mathbb{N}
\ee
and that the lattice description implies not only (\ref{r11s}) but also \cite{RP0707} 
\be
 \R_r^b=(r,1)\otimes\R_1^b,\qquad r\in\mathbb{N}
\label{r1R}
\ee
The fusion rules of the full Kac fusion algebra now follow straightforwardly using
the requirement of commutativity and associativity as we then have
\bea
 (r,b+kp)\otimes(r',b'+k'p)&=&\big((r,1)\otimes(r',1)\big)\otimes\big((1,b+kp)\otimes(1,b'+k'p)\big)\nn
 \R_r^b\otimes(r',b'+k'p)&=&\big((r,1)\otimes(r',1)\big)\otimes\big(\R_1^b\otimes(1,b'+k'p)\big)\nn
 \R_r^b\otimes\R_{r'}^{b'}&=&\big((r,1)\otimes(r',1)\big)\otimes\big(\R_1^b\otimes\R_1^{b'}\big)
\eea
The last of these relations is not needed to determine the full Kac fusion algebra
but must be satisfied for self-consistency of the fusion algebra.
The fusion rules needed to complete the Kac fusion algebra are
\bea
 (r,b+kp)\otimes(r',b'+k'p)&=&
   \bigoplus_{i=|r-r'|+1,\,\mathrm{by}\,2}^{r+r'-1}
   \bigg\{
   \bigoplus_{j=|k-k'|+1,\,\mathrm{by}\,2}^{k+k'-1}\ \,
   \bigoplus_{\ell=|i-j|+1,\,\mathrm{by}\,2}^{i+j-1}
   \bigoplus_{\beta}^{p-|b-b'|-1}
   \R_\ell^\beta\nn
 &\oplus&   
   \bigoplus_{j=|k-k'+\mathrm{sg}(b-b')|+1,\,\mathrm{by}\,2}^{k+k'}\ \,
   \bigoplus_{\ell=|i-j|+1,\,\mathrm{by}\,2}^{i+j-1}
   \bigoplus_{\beta}^{|b-b'|-1}  
   \R_\ell^\beta\nn
 &\oplus&
   \bigoplus_{\ell=|i-k-k'-1|+1,\,\mathrm{by}\,2}^{i+k+k'}
   \bigoplus_{\beta}^{b+b'-p-1}
   \R_\ell^\beta
   \oplus
   \bigoplus_{\beta=|b-b'|+1,\,\mathrm{by}\,2}^{p-|p-b-b'|-1}(i,\beta+(k+k')p)
   \bigg\}\nn
 \R_{r}^{b}\otimes(r',b'+k'p)&=&\bigg(
   \bigg\{
\bigoplus_{j=|r-k'-1|+1,\,\mathrm{by}\,2}^{r+k'-2}
  \!\!\bigoplus_{\beta}^{p-|p-b-b'|-1}
  \oplus
    \bigoplus_{j=|r-k'-1+\mathrm{sg}(p-b-b')|+1,\,\mathrm{by}\,2}^{r+k'-\mathrm{sg}(p-b-b')}
  \!\!\bigoplus_{\beta}^{|p-b-b'|-1} 
   \nn
 &\oplus&  \bigoplus_{j=|r-k'|+1,\,\mathrm{by}\,2}^{r+k'-1}
  \!\!\bigoplus_{\beta}^{p-|b-b'|-1}
  \oplus
  \bigoplus_{j=|r-k'+\mathrm{sg}(b-b')|+1,\,\mathrm{by}\,2}^{r+k'}
  \!\!\bigoplus_{\beta}^{|b-b'|-1}
  \bigg\}\bigoplus_{\ell=|r'-j|+1,\,\mathrm{by}\,2}^{r'+j-1}\R_{\ell}^\beta\nn
 &\oplus&
    \bigoplus_{\ell=|r-r'+k'|+1,\,\mathrm{by}\,2}^{r+r'+k'-1}
   \bigg\{
 \bigoplus_{\beta}^{b'-b-1}
  \oplus\bigoplus_{\beta=|b-b'|+1,\,\mathrm{by}\,2}^{p-|p-b-b'|-1}
 \bigg\}\R_{\ell}^\beta
  \bigg)/(1+\delta_{b,0})
 \eea
It follows, in particular, that the fundamental fusion algebra (\ref{fund}) is a subalgebra 
of the Kac fusion algebra (\ref{full}).
The fusion rules for critical dense polymers ${\cal LM}(1,2)$ are summarized in
Appendix~\ref{AppCrit}.

Recalling that $\R_r^0\equiv(1,rp)\equiv(r,p)$, it is noted that the modules
\be
 \big\{\R_r^b;\, r\in\mathbb{N},\, b\in\mathbb{Z}_{0,p-1}\big\}
\label{proj}
\ee
form an {\em ideal} of the Kac fusion algebra. This is in accordance with the
expectation that these modules are {\em projective}.

\subsubsection{Evidence for the fusion conjecture: lattice approach}
\label{SecEviFusionLattice}

The lattice approach to the logarithmic minimal model ${\cal LM}(1,p)$ 
\cite{PRZ0607,RP0707} is based on a loop model with loop fugacity
\be
 \beta=-2\cos\tfrac{\pi}{p}
\ee
Here we are interested in the model defined on strips of width $N$.
To describe the vertical Kac representations and their fusions, 
it suffices to consider the hamiltonian defined by
\be
 H=-\sum_{j=1}^{N-1}e_j
\ee
where $\{e_j;\ j\in\mathbb{Z}_{1,N-1}\}$ is the set of Temperley-Lieb generators
acting on $N$ strands.
Empirically \cite{PRZ0607}, the character of the Kac representation $(1,s)$ arises
in the scaling limit of the spectrum of the hamiltonian acting
on link states with exactly $s-1$ defects.
Viewing these defects as linked to the right (or left) boundary,
the Kac representation is associated with the corresponding boundary condition.
In particular, there are $N-1$ link states with exactly $N-2$ defects and our
choice of canonical ordering of these link states is 
\be
 \bigcap\,\big|\,\big|\,\ldots\,\big|\,\big|\,,\qquad 
 \big|\,\bigcap\,\big|\,\ldots\,\big|\,\big|\,,\qquad 
   \ldots\ldots,\qquad 
 \big|\,\big|\,\ldots\,\big|\,\big|\,\bigcap
\label{can}
\ee
We refer to \cite{PR0610} for more details.

Fusion is implemented diagrammatically by considering non-trivial
boundary conditions on {\em both} sides of the bulk.
In the diagrammatic description of the fusion product $(1,s)\otimes(1,s')$,
there are thus $s-1$ and $s'-1$ links emanating from the left and right boundaries,
respectively. As links from the left boundary can be joined with links from the right
boundary to form half-arcs above the bulk, the number of defects propagating through 
the bulk is given by $s+s'-2-2t$ where $0\leq t\leq \min\{s,s'\}-1$.
In the last expression in (\ref{chardec}), the integer $t$ labels the number of such half-arcs
linking the two boundaries. For given $t$, we thus have $s-t-1$ and $s'-t-1$
half-arcs linking the bulk to the left and right boundary, respectively.

As usual, we group the link states according to their number of half-arcs linking
the bulk to the boundaries and order these groups with increasing such numbers.
The resulting matrix representation of the hamiltonian is then upper block-triangular
with vanishing blocks beyond the first super-diagonal of blocks.
It is recalled that we do not anticipate Jordan cells of ranks greater than 2
in the hamiltonian.
To examine Jordan cells of rank 2 formed
between {\em neighbouring} blocks on the diagonal, 
it thus suffices to analyze the upper block-triangular
matrix defined by the four adjacent blocks spanned diagonally by the said two blocks.
This gives insight into the appearance of rank-2 modules of the type $\R_r^1$.
Beyond neighbouring blocks, care has to be taken, though, since some
non-trivial Jordan cells are formed using `ligatures', see (\ref{M}) below.
This is indeed the case for $\R_r^b$ for $b>1$ since such a rank-2 module
can be viewed as an indecomposable sum (\ref{Rrb})
of two Kac representations corresponding to boundary conditions differing in numbers
of defects by $2b>2$. The responsible Jordan cells are thus formed
between blocks which are {\em not} neighbours. 

As illustration of this `ligature phenomenon', we consider the matrix
\be
 M=\begin{pmatrix}a&1&0\\ 0&b&1\\ 0&0&a\end{pmatrix}
\label{M}
\ee
For $a\neq b$, its Jordan canonical form reads
\be
 J=S^{-1}MS=\begin{pmatrix}b&0&0\\0&a&1\\0&0&a\end{pmatrix}
\ee
where
\be
 S^{-1}=\begin{pmatrix}0&-\sigma&1\\ \sigma&1&0\\ 0&0&1\end{pmatrix},\hspace{1cm}
  S=\begin{pmatrix}\sigma^{-2}&\sigma^{-1}&-\sigma^{-2}\\ -\sigma^{-1}&0&\sigma^{-1}\\
    0&0&1\end{pmatrix},\hspace{1cm}\sigma=a-b
\ee
That is, a rank-2 Jordan cell is formed between the two copies of the degenerate
eigenvalue $a$.
If, on the other hand, we eliminate the second row and column from $M$ {\em before}
examining the possibility of a rank-2 Jordan cell, we end up with the {\em diagonal} matrix
${\rm diag}(a,a)$. A search for non-trivial Jordan cells can therefore not be conducted this
naively, and focus here is on neighbouring blocks. That is, we are only concerned with the
appearance of rank-2 modules of the type $\R_r^1$. 
It is also noted that permutations alone cannot resolve the indicated problems associated with treating
blocks which are not neighbours. This is again illustrated by the matrix $M$ in (\ref{M})
which is similar to
\be
  P^{-1}MP=\begin{pmatrix}a&0&1\\ 0&a&0\\ 0&1&b\end{pmatrix},\qquad
  P=\begin{pmatrix}1&0&0\\ 0&0&1\\ 0&1&0\end{pmatrix}
\ee
However, the matrix $P^{-1}MP$ is not upper block-triangular with vanishing blocks beyond the 
first super-diagonal of blocks.

Now, let us implement the fusion product $(1,s)\otimes(1,s')$ for $s,s'>1$
on a lattice of limited system size 
\be
 N=s+s'-2-2t,\qquad t=0,1,\ldots,\min\{s,s'\}-1
\ee 
for some $t$ in the range given.
This means that the bulk can accommodate up to $N$ defects while there must
be at least $t$ half-arcs linking the two boundaries.
In the decomposition (\ref{ss}), the $t$ rightmost Kac representations are therefore not
present while the remaining ones are
\be
 (1,s)\otimes(1,s')\big|_{N}
   =(1,|s-s'|+1)\stackrel{\mbox{?}}{\oplus}
   \ldots\stackrel{\mbox{?}}{\oplus}(1,s+s'-3-2t)\stackrel{\mbox{?}}{\oplus}(1,s+s'-1-2t)
\label{ssN}
\ee 
To gain insight into whether the final sum in this decomposition is
{\em direct} or {\em indecomposable}, 
we will now characterize when a non-trivial Jordan cell is formed in the
hamiltonian $H_{s,s'}^{(N)}$ between the two neighbouring blocks corresponding to 
$N-2$ or $N$ defects, respectively.
For $N=6$, using the ordered basis (\ref{can}), 
the corresponding matrix realization of the hamiltonian is given by
\be
 -H_{s,s'}^{(6)}=\left(\!\!\begin{array}{rccccl}
  \beta&1&0&0&0&\delta_{s-t,2}\\[.2cm]
  1&\beta&1&0&0&\delta_{s-t,3}\\[.2cm]
  0&1&\beta&1&0&\delta_{s-t,4}\\[.2cm]
  0&0&1&\beta&1&\delta_{s-t,5}\\[.2cm]
  0&0&0&1&\beta&\delta_{s-t,6}\\[.2cm]
  0&0&0&0&0&0
  \end{array}\!\!\!\right)
\ee
The extension to general $N$ is straightforward and discussed in
Appendix~\ref{AppJordan}. We thus find that
$H_{s,s'}^{(N)}$ is diagonalizable unless
there exists $j_0\in\mathbb{Z}_{1,N-1}$ for which
$\beta+2\cos\frac{j_0\pi}{N}=0$ and $\sin\frac{j_0(s-t-1)\pi}{N}\neq0$ in which case the 
Jordan canonical form of $H_{s,s'}^{(N)}$ contains a single non-trivial Jordan cell. 
This cell is of rank 2 and has diagonal elements 0.
It follows that this non-trivial Jordan cell appears if and only if
\be
 p\mid (s+s'-2-2t),\qquad p\nmid (s-t-1)
\label{pss}
\ee
Since the pair of conditions $q\,|\,(n+m)$ and $q\nmid n$ implies $q\nmid m$, we may restore
the symmetry between $s$ and $s'$ in (\ref{pss}) by redundantly including $p\nmid (s'-t-1)$.
This symmetry is a manifestation of the commutativity of the fusion product $(1,s)\otimes(1,s')$,
of the equivalence of the left- and right-sided decompositions of the diagrammatic
implementation of this fusion product, and of the choice of canonical ordering of the link states
with $N-2$ defects (\ref{can}).

This {\em exact} result (\ref{pss}) for finite system sizes is in accordance with the fusion rule
(\ref{fullver2}) for $(1,s)\otimes(1,s')$. Indeed, assuming that the observed 
Jordan-cell structures survive in the
continuum scaling limit, the result provides valuable insight used to determine 
whether the particular sum 
\be
 (1,s)\otimes(1,s')=\ldots (1,s+s'-3-2t)\stackrel{\mbox{?}}{\oplus}(1,s+s'-1-2t)\ldots
\ee
in the decomposition (\ref{ss}) of the fusion product is {\em direct} or {\em indecomposable}. 
{}From the lattice analysis above, we thus conclude that it is indecomposable due
to the presence of non-trivial Jordan cells if and only if the conditions in (\ref{pss}) are satisfied.
For this to be compatible with the conjectured fusion rules, the latter must predict
that the rank-2 module $\R_r^1$ appears (with multiplicity 1) in the decomposition
of $(1,s)\otimes(1,s')$ if and only if 
\be
 \exists\, \tau\in\mathbb{Z}_{1,\min\{s,s'\}-1}:\qquad rp=|s-s'|+2\tau,\qquad \tau\not\in p\mathbb{N}
\ee
Writing $\tau=a+\ell p$, this is easily verified.

{}From the lattice approach, we now know where certain Jordan cells appear
in the decomposition of $(1,s)\otimes(1,s')$,
but in general, this is not sufficient to determine the various representations.
In critical dense polymers ${\cal LM}(1,2)$, for example, we have thus found that
\be
 (1,3)\otimes(1,3)=(1,1)\oplus_i(1,3)\stackrel{\mbox{?}}{\oplus}(1,5)
\label{1311}
\ee
where the indecomposable sum is due to the formation of non-trivial Jordan cells.
The lattice approach offers an additional clue.
Continuing the examination of (\ref{1311}), we note that the link states
associated with the subfactor $(1,1)$ (corresponding to $t=2$) and the link states
associated with the subfactor $(1,3)$ (corresponding to $t=1$) all contain a half-arc
linking the two boundaries. Ignoring this common spectator half-arc, the diagrammatic
description becomes equivalent to the lattice implementation of the fusion product
\be
 (1,2)\otimes(1,2)=\R_1^1
\label{1212}
\ee
Alternatively, we may focus on the link states associated with the subfactors
$(1,3)$ and $(1,5)$ corresponding to $t=1$ or $t=0$, respectively.
Unlike before, this does not correspond to a single fusion product.
The only candidate with the same number of defects propagating through the bulk
is $(1,2)\otimes(1,4)$, but this is associated with link states with 1 and 3 links emanating 
from the left and right boundaries, respectively.
We thus conclude that the fusion product $(1,3)\otimes(1,3)$ contains
the rank-2 module $\R_1^1$ as a subfactor, that is,
\be
 (1,3)\otimes(1,3)=\R_1^1\stackrel{\mbox{?}}{\oplus}(1,5)
\label{131315}
\ee
Below, we supplement this lattice analysis of the fusion product $(1,3)\otimes(1,3)$
by applications of the NGK algorithm.

\subsubsection{Evidence for the fusion conjecture: NGK algorithm}
\label{SecEviFusionNGK}

A priori, the right side of (\ref{131315}) could correspond to a single indecomposable 
representation (since it remains to be established that (\ref{proj}) is the set of projective 
representations). 
According to the conjectured fusion rules (\ref{fullver2}), however, the full decomposition
reads
\be
 (1,3)\otimes(1,3)=\R_1^1\oplus(1,5)
\label{1313R15}
\ee
To test this, we have applied the NGK algorithm to the fusion product $(1,3)\otimes(1,3)$,
assuming that $(1,3)$ is a highest-weight module. 
Details of this analysis to Nahm level 2 appear in Appendix~\ref{App1313}, and they confirm
the fusion rule (\ref{1313R15}). They also confirm that the Kac representation $(1,5)$ is
a highest-weight module and not its contragredient module.
Likewise in ${\cal LM}(1,2)$, we have confirmed the fusion rule
\be
 (1,3)\otimes(1,5)=\R_2^1\oplus(1,7)
\ee
and the highest-weight property of $(1,7)$ to Nahm level 3.

As observed in Section~\ref{SecRing} below,
the vertical Kac representations $(1,s)$ are all generated from repeated fusion
of $(1,2)$ and $(1,p+1)$. In accordance with the results of the NGK algorithm,
it is therefore natural to expect that the Kac representations $(1,s)$ thereby generated
are all highest-weight modules provided that $(1,p+1)$ is. 
It thus {\em suffices} to assume that $(1,p+1)$ is a highest-weight module.
\\[.2cm]
\noindent
{\bf Refined highest-weight assumption.}\quad (i) The Kac representation $(1,p+1)$
is a highest-weight Virasoro module. (ii) Repeated fusion subsequently ensures that
all vertical Kac representations $(1,s)$ are highest-weight Virasoro modules.
\\[.2cm]
Our analysis does not, however, provide direct arguments for the assumption that
the Kac representation $(1,p+1)$ is a highest-weight module.
As we will see below, the fusion rules actually turn out to be independent of 
whether $(1,p+1)$ is indeed a highest-weight module or the corresponding
contragredient module.

\subsubsection{Even and odd sectors}

{}From the lattice approach, it is of interest to understand the continuum scaling limit of the
situation where the only constraint on the number
of defects is that it is of the same parity as the bulk system size $N$. Depending on the parity of 
$N$, we refer to the two possible scenarios as the {\em even} and {\em odd sectors}.
They can be viewed as systems with {\em free boundary conditions}, but they can also
be interpreted as finitized versions of the fusion products
\be
 (1,\tfrac{N+2}{2})\otimes(1,\tfrac{N+2}{2})\quad\mathrm{and}\quad
 (1,\tfrac{N+1}{2})\otimes(1,\tfrac{N+3}{2})
\label{1N1N}
\ee
respectively. To examine the continuum scaling limit of a system with free boundary conditions,
we can thus resort to the fusion rules for the fusions in (\ref{1N1N}) as given in (\ref{fullver2}).
For $b\in\mathbb{Z}_{0,p-1}$ and $k\in\mathbb{N}_0$, the first fusion rule in 
(\ref{fullver2}) yields
\be
\begin{array}{rcl}
 (1,b+kp)\otimes(1,b+kp)&=&
 \displaystyle{
  \bigoplus_{j=1}^k\,\bigoplus_\beta^{p-1}\R_{2j-1}^\beta
  \oplus\bigoplus_\beta^{2b-p-1}\R_{2k+1}^\beta
  \oplus\bigoplus_\beta^{p-|p-2b|-1}(1,\beta+2kp)
 }\\[.7cm]
 (1,b+kp)\otimes(1,b+1+kp)&=&
 \displaystyle{
  \bigoplus_{j=1}^k\,\bigoplus_\beta^{p-2}\R_{2j-1}^\beta
  \oplus\bigoplus_\beta^{2b-p}\R_{2k+1}^\beta
  \oplus\bigoplus_{j=1}^k\R_{2j}^0
  \oplus\bigoplus_{\beta=2,\,\mathrm{by}\,2}^{p-|p-2b-1|-1}(1,\beta+2kp)
 }
\end{array}
\ee
It is verified that the second of these rules applies for $b=p-1$, even though $b'=p$ in that case.
It follows that the continuum scaling limit of a system with free boundary conditions 
is described by
\be
 \lim_{n\to\infty}(1,n)\otimes(1,n)=\bigoplus_{j\in\mathbb{N}}\,
  \bigoplus_\beta^{p-1}\R_{2j-1}^\beta,\qquad
 \lim_{n\to\infty}(1,n)\otimes(1,n+1)=\bigoplus_{j\in\mathbb{N}}\Big(
  \R_{2j}^0\oplus\bigoplus_\beta^{p-2}\R_{2j-1}^\beta\Big)
\ee
in accordance with the recent analysis of Jordan structures in~\cite{MS1101}.
In particular, for critical dense polymers as described by ${\cal LM}(1,2)$~\cite{PR0610}, 
we thus have
\be
 \lim_{n\to\infty}(1,n)\otimes(1,n)=\bigoplus_{j\in\mathbb{N}}\R_{2j-1}^1,\qquad
 \lim_{n\to\infty}(1,n)\otimes(1,n+1)=\bigoplus_{j\in\mathbb{N}}\,(1,2j)
\ee
showing that reducible yet indecomposable representations only arise in the even sector.

\subsection{Contragredient extension}
\label{SecContraFusion}

It is stressed that the set
\be
 \Jc^{\mathrm{Kac}}=
  \big\{(r,s),\R_r^b;\,r,s\in\mathbb{N},\,b\in\mathbb{Z}_{1,p-1}\big\}
\ee
of representations appearing in the Kac fusion algebra exhausts the set of representations
associated with boundary conditions in \cite{PRZ0607,RP0707}.
Extending this set by the contragredient Kac representations
\be
 \Jc^{\mathrm{Kac}}\,\to\,
 \Jc^{\mathrm{Cont}}=\Jc^{\mathrm{Kac}}\cup
  \big\{(r,s)^\ast;\,r,s\in\mathbb{N}\big\}
\label{sets}
\ee
gives rise to the larger fusion algebra
\be
 \big\langle\Jc^{\mathrm{Cont}}\big\rangle
 =\big\langle(r,s),(r,s)^\ast,\R_r^b;\, r,s\in\mathbb{N},\,b\in\mathbb{Z}_{1,p-1}\big\rangle
\label{FusCon}
\ee
where we recall (\ref{rsast}).
A priori, additional representations could be generated by repeated fusion
of the representations listed.
However, preliminary evaluations of a variety of fusion products 
seem to suggest that the extended fusion algebra (\ref{FusCon})
closes on the set of representations listed.
To describe this fusion algebra, we introduce
\be
 \Cc_n[(r,s)]=\begin{cases} 
  (r,s),\quad &n>0
  \\[.2cm]
  (r,s)^\ast,\quad &n<0
 \end{cases}
\label{C}
\ee
In our applications, $\Cc_0[(r,s)]$ only appears if $(r,s)$ is irreducible
in which case
\be
 \Cc_0[(r,s)]=(r,s)=(r,s)^\ast,\qquad s\in\mathbb{Z}_{1,p-1}\cup p\mathbb{N}
\ee
where we have extended the definition of $\Cc_0$ to all fully reducible representations 
(\ref{rsast}).
\\[.2cm]
{\bf Contragredient fusion conjecture.}\quad 
The fusion rules involving contragredient Kac representations in 
the extended fusion algebra (\ref{FusCon}) are given by or follow readily from
\be
 (r,s)^\ast\otimes(r',s')^\ast=\big((r,s)\otimes(r',s')\big)^\ast,\qquad
 \R_r^b\otimes(r',s')^\ast=\R_r^b\otimes(r',s')
\label{rsrs}
\ee
and
\bea
 (1,b+kp)\otimes(1,b'+k'p)^\ast&=&\!\!\bigoplus_{j=|k-k'|+2,\,\mathrm{by}\,2}^{k+k'}\!\!\!\!
   \bigoplus_{\beta}^{p-|p-b-b'|-1}\!\!\R_j^\beta\oplus
   \bigoplus_{j=|k-k'|+1,\,\mathrm{by}\,2}^{k+k'-\mathrm{sg}(p-b-b')}\
   \bigoplus_{\beta}^{|p-b-b'|-1}\!\!\R_j^\beta  
\nn
&\oplus&\!\!\bigoplus_\beta^{(b-b')\mathrm{sg}(k'-k)-1}\!\R_{|k-k'|}^\beta
 \oplus\bigoplus_{\beta=|b-b'|+1,\,\mathrm{by}\,2}^{p-|p-b-b'|-1}
 \!\!\Cc_{k-k'}[(1,\beta+|k-k'|p)]
\label{bkbk}
\eea
where $b,b'\in\mathbb{Z}_{0,p-1}$ and $k,k'\in\mathbb{N}_{0}$. 
\\[.2cm]
Since $(r,1)$ is irreducible, we thus have
\be
 (r,s)^\ast=(r,1)^\ast\otimes(1,s)^\ast=(r,1)\otimes(1,s)^\ast
\ee
from which it follows that the general fusion product $(r,s)\otimes(r',s')^\ast$ can be
computed as
\be
 (r,s)\otimes(r',s')^\ast=\big((r,1)\otimes(r',1)\big)\otimes\big((1,s)\otimes(1,s')^\ast\big)
\ee
This yields the general fusion rule
\bea
 (r,b+kp)\otimes(r',b'+k'p)^\ast&=&
    \bigoplus_{i=|r-r'|+1,\,\mathrm{by}\,2}^{r+r'-1}
   \bigg\{
   \bigoplus_{j=|k-k'|+2,\,\mathrm{by}\,2}^{k+k'}\ \,
   \bigoplus_{\ell=|i-j|+1,\,\mathrm{by}\,2}^{i+j-1}
   \bigoplus_{\beta}^{p-|p-b-b'|-1}
   \R_\ell^\beta\nn
 &\oplus&   
   \bigoplus_{j=|k-k'|+1,\,\mathrm{by}\,2}^{k+k'-\mathrm{sg}(p-b-b')}\ \,
   \bigoplus_{\ell=|i-j|+1,\,\mathrm{by}\,2}^{i+j-1}
   \bigoplus_{\beta}^{|p-b-b'|-1}  
   \R_\ell^\beta\nn
 &\oplus&
   \bigoplus_{\ell=|i-|k-k'||+1,\,\mathrm{by}\,2}^{i+|k-k'|-1}
   \bigoplus_{\beta}^{(b-b')\mathrm{sg}(k'-k)-1}
   \R_\ell^\beta\nn
 &\oplus&
   \bigoplus_{\beta=|b-b'|+1,\,\mathrm{by}\,2}^{p-|p-b-b'|-1}\!\!\Cc_{k-k'}[(i,\beta+|k-k'|p)]
   \bigg\}
\label{bkbk2}
\eea
In general, the fusion rules are not invariant under replacement by contragredient modules
as illustrated by
\be
 (1,1)^\ast\otimes(r,s)=(r,s)\neq(r,s)^\ast=(1,1)\otimes(r,s)^\ast,\qquad p<s\neq kp
\ee
These trivial fusion rules are encoded in (\ref{bkbk2}) and correspond to 
$r'=1,b'=1,k'=0$ or $r=1,b=1,k=0$, respectively.
As a consequence of (\ref{rsrs}), 
we note that the extended fusion algebra contains the two isomorphic fusion subalgebras
\be
 \big\langle(r,s),\R_r^b;\,r,s\in\mathbb{N},\,b\in\mathbb{Z}_{1,p-1}\big\rangle
 \simeq 
 \big\langle(r,s)^\ast,\R_r^b;\,r,s\in\mathbb{N},\,b\in\mathbb{Z}_{1,p-1}\big\rangle
\label{KacKac}
\ee
of which the first one is the Kac fusion algebra.
It is also noted that the representations in (\ref{proj}) form an {\em ideal} of the
extended fusion algebra, still in accordance with the representations being projective.

The fusion rules (\ref{bkbk}) can be obtained by extending the applications of the
forgetful functor (\ref{F}) with 
\be
 \Fc[(1,s)^\ast]=(1,s)
\ee
and subsequently modifying the prescription or algorithm discussed following (\ref{Ffus}).
In that discussion, we formed rank-2 modules starting with the {\em lowest} Kac label
-- now we start with the {\em greatest} Kac label.
That is, after applying the forgetful functor to the fusion product $(1,s)\otimes(1,s')^\ast$
\be
 \Fc[(1,s)\otimes(1,s')^\ast]=\bigoplus_{s''=|s-s'|+1,\,\mathrm{by}\,2}^{s+s'-1}(1,s'')
\label{Ffusast}
\ee
we initially focus on the Kac representation $(1,s_1'')$ with maximal Kac
label $s_1''=s+s'-1$. Depending on $p$, this will appear as the submodule $(1,rp+b)$ of
the rank-2 module $\R_r^b$ if and only if the matching module $(1,rp-b)$ also appears in the
decomposition in (\ref{Ffusast}). If not, the (contragredient) Kac representation 
$\Cc_{s-s'}[(1,s_1'')]$
will appear `by itself' in the decomposition of the fusion product. 
If a rank-2 module is not formed for $s=s'$,
the two options $(1,s_1'')$ and $(1,s_1'')^\ast$ turn out to be identical.
Having completed the examination of $(1,s_1'')$, we remove it
together with its potential partner $(1,rp-b)$ from the direct sum
in (\ref{Ffusast}) and repeat the analysis for $(1,s_2'')$ corresponding to the new maximal 
Kac label $s_2''$. As before, this algorithm is continued until all the Kac representations 
in (\ref{Ffusast}) have been accounted for. 
It is straightforward to verify that this prescription yields the fusion rules (\ref{bkbk}).

\subsection{Polynomial fusion rings}
\label{SecRing}

Together with the fact that the fundamental fusion algebra is a subalgebra of the 
Kac fusion algebra, the fusion rules
\bea
 (1,2)\otimes(1,kp+b)&=&(1,kp+b-1)\oplus(1,kp+b+1)\nn
 (1,p+1)\otimes(1,kp+b)&=&\bigoplus_{\beta}^{p-b}\R_k^\beta
  \oplus\bigoplus_\beta^{b-2}\R_{k+1}^\beta
  \oplus(1,(k+1)p+b)
\label{12p}
\eea
demonstrate that the Kac fusion algebra
is generated from repeated fusion of the Kac representations
\be
 \big\{(1,1),(2,1),(1,2),(1,p+1)\big\}
\ee
that is,
\be
 \big\langle\Jc^{\mathrm{Kac}}\big\rangle
 =\big\langle(r,s);\ r,s\in\mathbb{N}\big\rangle
 =\big\langle(1,1),(2,1),(1,2),(1,p+1)\big\rangle
\label{full2}
\ee
It is therefore natural to expect that this fusion algebra is
isomorphic to a polynomial ring in the three entities
$X\leftrightarrow(2,1)$, $Y\leftrightarrow(1,2)$ and $Z\leftrightarrow(1,p+1)$. 
This is indeed what we find.
\\[.2cm]
{\bf Proposition 1.}\quad The Kac fusion algebra is isomorphic to the polynomial ring generated
by $X$, $Y$ and $Z$ modulo the ideal $(P_p(X,Y),Q_p(Y,Z))$, that is,
\be
 \big\langle\Jc^{\mathrm{Kac}}\big\rangle
  \simeq\mathbb{C}[X,Y,Z]/\big(P_p(X,Y),Q_p(Y,Z)\big)
\ee
where
\be
 P_p(X,Y)=\big[X-2T_p(\tfrac{Y}{2})\big]U_{p-1}(\tfrac{Y}{2}),\qquad
 Q_p(Y,Z)=\big[Z-U_p(\tfrac{Y}{2})\big]U_{p-1}(\tfrac{Y}{2})
\label{PQ}
\ee
For $r\in\mathbb{N}$, $k\in\mathbb{N}_0$ and $b\in\mathbb{Z}_{0,p-1}$, 
the isomorphism reads
\bea
 (r,kp+b)&\leftrightarrow& 
   U_{r-1}(\tfrac{X}{2})
    \Big(U_{kp+b-1}(\tfrac{Y}{2})+\big[Z^k-U_{p}^k(\tfrac{Y}{2})\big]U_{b-1}(\tfrac{Y}{2})\Big)\nn
 \R_r^b&\leftrightarrow&
   (2-\delta_{b,0})U_{r-1}(\tfrac{X}{2})T_b(\tfrac{Y}{2})U_{p-1}(\tfrac{Y}{2})
\eea
where $T_n(x)$ and $U_n(x)$ are Chebyshev polynomials of the first and second kind, 
respectively.
\\[.2cm]
{\bf Proof.}\quad The relation $P_p(X,Y)=0$ corresponds to the identification
$(2,p)\equiv(1,2p)$ and encodes $(r,p)\equiv(1,rp)$ more generally, cf.\! (\ref{UU}), 
while the relation $Q_p(Y,Z)=0$ follows from the fusion rule
\be
 (1,p)\otimes(1,p+1)=\bigoplus_{\beta}^{p-2}\R_1^\beta\oplus(1,2p)
\ee
The remaining fusion rules are then verified straightforwardly in the polynomial ring.
Here we only demonstrate the two fusion rules in (\ref{12p}).
The first of these follows immediately from the recursion relation for the Chebyshev
polynomials. To show the second of the fusion rules, we note
the basic decomposition rules
\be
 U_m(x)U_n(x)=\sum_{j=|m-n|,\,\mathrm{by}\,2}^{m+n}U_j(x),\qquad
 2T_m(x)U_{n-1}(x)=U_{n+m-1}(x)+\mathrm{sg}(n-m)U_{|n-m|-1}(x)
\ee
where $U_{-1}(x)=0$. As a consequence, we have
\be
 U_{p-1}(x)\sum_{j=0}^{k-1}U_p^{k-j-1}(x)U_{jp+b-2}(x)=U_{b-1}(x)U_p^k(x)-U_{kp+b-1}(x)
\ee
which is established by induction in $k$ and shows that the expression on the right side is
divisible by $U_{p-1}(x)$. This is of importance when multiplied by $Z$ due to
the form of $Q_p(Y,Z)$. With the additional observation that
\be
 U_{r-1}(\tfrac{X}{2})U_{p-1}(\tfrac{Y}{2})\equiv U_{rp-1}(\tfrac{Y}{2})\quad 
   (\mathrm{mod}\ P_p(X,Y))
\label{UU}
\ee
which follows by induction in $r$, the second fusion rule readily follows.
$\quad\Box$
\\[.2cm]
Extending the arguments just presented for the Kac fusion algebra, one
finds that the extended Kac fusion algebra 
(\ref{FusCon}) is also generated from repeated fusion of a small number of Kac representations
\be
 \big\langle(r,s),(r,s)^\ast;\ r,s\in\mathbb{N}\big\rangle
  =\big\langle(1,1),(2,1),(1,2),(1,p+1),(1,p+1)^\ast\big\rangle
\label{full3}
\ee
and that it is isomorphic to a polynomial ring.
\\[.2cm]
\noindent
{\bf Proposition 2.}\quad The extended Kac fusion algebra (\ref{FusCon})
is isomorphic to the 
polynomial ring generated
by $X$, $Y$, $Z$ and $Z^\ast$ modulo the ideal 
$(P_p(X,Y),Q_p(Y,Z),Q_p(Y,Z^\ast),R_p(Y,Z,Z^\ast))$, that is,
\be
 \big\langle\Jc^{\mathrm{Cont}}\big\rangle
  \simeq\mathbb{C}[X,Y,Z,Z^\ast]/\big(P_p(X,Y),Q_p(Y,Z),Q_p(Y,Z^\ast),R_p(Y,Z,Z^\ast)\big)
\ee
where the polynomials $P_p$ and $Q_p$ are defined in (\ref{PQ}) while
\be
 R_p(Y,Z,Z^\ast)=ZZ^\ast-U_p^2(\tfrac{Y}{2})
\label{R}
\ee
For $r\in\mathbb{N}$, $k\in\mathbb{N}_0$ and $b\in\mathbb{Z}_{0,p-1}$, 
the isomorphism reads
\bea
 (r,kp+b)&\leftrightarrow& 
   U_{r-1}(\tfrac{X}{2})
    \Big(U_{kp+b-1}(\tfrac{Y}{2})+\big[Z^k-U_{p}^k(\tfrac{Y}{2})\big]U_{b-1}(\tfrac{Y}{2})\Big)\nn
 (r,kp+b)^\ast&\leftrightarrow& 
   U_{r-1}(\tfrac{X}{2})
    \Big(U_{kp+b-1}(\tfrac{Y}{2})+\big[(Z^\ast)^k-U_{p}^k(\tfrac{Y}{2})\big]U_{b-1}(\tfrac{Y}{2})\Big)\nn
 \R_r^b&\leftrightarrow&
   (2-\delta_{b,0})U_{r-1}(\tfrac{X}{2})T_b(\tfrac{Y}{2})U_{p-1}(\tfrac{Y}{2})
\eea
{\bf Proof.}\quad Compared to the proof of Proposition 1, the essential new feature
is the appearance of $Z^\ast$. The relation $Q_p(Y,Z^\ast)=0$ plays the same
role for the contragredient Kac representations and $Z^\ast$ as
$Q_p(Y,Z)=0$ does for the Kac representations and $Z$. This yields the
part of the polynomial ring corresponding to (\ref{KacKac}).
The relation $R_p(Y,Z,Z^\ast)=0$ corresponds to the fusion rule
\be
 (1,p+1)\otimes(1,p+1)^\ast=(1,1)\oplus\bigoplus_\beta^{p-3}\R_1^\beta\oplus\R_2^1
\ee
To establish the general fusion rule (\ref{bkbk}) in the ring picture, we first use induction in $n$
to establish
\be
 U_p^{2n}(\tfrac{Y}{2})Z^m\equiv Z^m
   +\sum_{j=0}^{n-1}U_p^{m+2j}(\tfrac{Y}{2})U_{p-1}(\tfrac{Y}{2})
   U_{p+1}(\tfrac{Y}{2})
   \quad (\mathrm{mod}\ Q_p(Y,Z)),\qquad n\in\mathbb{N}
\ee
and similarly for $Z$ replaced by $Z^\ast$. This is needed when reducing
\be
 Z^k(Z^\ast)^{k'}\equiv U_p^{2\min(k,k')}(\tfrac{Y}{2})\begin{cases} Z^{k-k'},\quad&k\geq k' \\ 
   (Z^\ast)^{k'-k},\quad&k<k'  \end{cases}
   \qquad  (\mathrm{mod}\ R_p(Y,Z,Z^\ast))
\ee
For simplicity, we let $k\geq k'$ in which case we find
\be
 (1,b+kp)\otimes(1,b'+k'p)^\ast\leftrightarrow
  \big[Z^{k-k'}-U_p^{k-k'}(\tfrac{Y}{2})\big]U_{b-1}(\tfrac{Y}{2})U_{b'-1}(\tfrac{Y}{2})
    +U_{kp+b-1}(\tfrac{Y}{2})U_{k'p+b'-1}(\tfrac{Y}{2})
\ee
This polynomial expression is recognized as corresponding to the right side of (\ref{bkbk}).
$\quad\Box$

\section{Conclusion}

We have discussed the representation content and fusion algebras of the
logarithmic minimal model ${\cal LM}(1,p)$.
We have thus proposed a classification of the entire family of Kac representations 
as submodules of Feigin-Fuchs modules and presented a conjecture
for their fusion algebra. To test these proposals, we have used a combination
of the lattice approach to ${\cal LM}(1,p)$ and applications of the NGK algorithm.
We have also discussed a natural extension of the representation content by
inclusion of the modules contragredient to the Kac representations,
and we have presented a conjecture for the corresponding fusion algebra. 
This extended fusion algebra as well as the conjecture for the Kac fusion algebra itself
were then shown to be isomorphic to 
polynomial fusion rings which were described explicitly.

Continuing the work in \cite{BFGT0901} on a Kazhdan-Lusztig-dual
quantum group for the logarithmic minimal model ${\cal LM}(1,p)$,
fusion of Kac representations is considered in \cite{BGTnotes}.
The corresponding fusion algebra appears to be equivalent to the one
discussed here. This is a very reassuring observation for both methodologies
and offers independent evidence for the Kac fusion algebra discussed here.

The work presented here pertains to the logarithmic minimal models
${\cal LM}(1,p)$, but the methods used in obtaining the various results
are expected to extend straightforwardly to the general family of
logarithmic minimal models ${\cal LM}(p,p')$.
We hope to discuss the corresponding classification of Kac 
representations and their fusion algebras elsewhere.
The case ${\cal LM}(2,3)$ is particularly interesting as it describes
critical percolation.

We find  the remarkably simple expression (\ref{D21})
in the singular vector conjecture very intriguing.
Preliminary results indicate that it can be extended from $\D_{2,1}$
to general $\D_{r,1}$ and even to general logarithmic minimal models ${\cal LM}(p,p')$.
We also hope to discuss this elsewhere.

In the ${\cal W}$-extended picture ${\cal WLM}(1,p)$, Yang-Baxter integrable boundary conditions
associated with irreducible or projective representations of the triplet ${\cal W}$-algebra
${\cal W}(p)$ were introduced in \cite{PRR0803}.
With the results of the present work, it is natural to expect that there also exist Yang-Baxter
integrable boundary conditions associated with the reducible yet indecomposable
${\cal W}(p)$-representations of rank 1 appearing in \cite{FGST0512}.
This is indeed what we find as we will discuss elsewhere \cite{Ras1106}.

\section*{Acknowledgments}
\vskip.1cm
\noindent
This work is supported by the Australian Research Council
under the Future Fellowship scheme, project number FT100100774. 
The author thanks Paul A. Pearce, David Ridout,
Philippe Ruelle and Ilya Yu.\! Tipunin 
for helpful discussions and comments,
and the authors of \cite{BGTnotes} for sharing their results prior to publication.

\appendix

\section{Nahm-Gaberdiel-Kausch algorithm}
\label{AppNGK}

Here we summarize some of the ingredients in the NGK algorithm,
but refer to the original papers \cite{Nahm9402,GK9604} as well as
\cite{EF0604,MR0708} for more details.

\subsection{Co-multiplication}

We are interested in the co-multiplications given by
\be
 \D(L_n)=\sum_{m=-1}^n\bin{n+1}{m+1}L_m\times I+I\times L_n,\qquad\quad 
   n\in\mathbb{Z}_{\geq-1}
\label{copos}
\ee
and
\bea
 \D(L_{-n})=\sum_{m=-1}^\infty(-1)^{m+1}\bin{n+m-1}{m+1}L_m\times I+I\times L_{-n},
   \qquad\quad n\in\mathbb{Z}_{\geq2}\nn
 \Dti(L_{-n})=L_{-n}\times I+(-1)^{n+1}\sum_{m=-1}^\infty\bin{n+m-1}{n-2}I\times L_m,
   \qquad\quad n\in\mathbb{Z}_{\geq2}
\label{coneg}
\eea
This should not be confused with the notation for conformal weights.
Useful examples of the co-multiplications are
\bea
 \D(L_2)&=&L_{-1}\times I+3 L_0\times I+3 L_1\times I+L_2\times I+I\times L_2\nn
 \D(L_1)&=&L_{-1}\times I+2L_0\times I+L_1\times I+I\times L_1\nn
 \D(L_0)&=&L_{-1}\times I+L_0\times I+I\times L_0\nn
 \D(L_{-1})&=&L_{-1}\times I+I\times L_{-1}\nn
 \D(L_{-2})&=&\big(\ldots-L_2\times I+L_1\times I-L_0\times I+L_{-1}\times I\big)+I\times L_{-2}\nn
 \Dti(L_{-2})&=&L_{-2}\times I-\big(\ldots+I\times L_2+I\times L_1+I\times L_0+I\times L_{-1}\big)
\eea

\subsection{Singular vectors}
\label{AppSing}

We denote the singular vector appearing at level $rs$ in the highest-weight Verma 
module $V_{r,s}$ by $\ket{\la_{r,s}}$ and normalize it by setting the coefficient to 
$L_{-1}^{rs}$ equal to 1. For $r=1,2,3,4,5$, the singular vector
$\ket{\la_{r,1}}$ is given by
\bea
 \ket{\la_{1,1}}&=&\big\{L_{-1}\big\}\ket{\D_{1,1}}\nn
 \ket{\la_{2,1}}&=&\big\{L_{-1}^2-pL_{-2}\big\}\ket{\D_{2,1}}\nn
 \ket{\la_{3,1}}&=&\big\{L_{-1}^3-4pL_{-2}L_{-1}+2p(2p-1)L_{-3}\big\}\ket{\D_{3,1}}\nn
 \ket{\la_{4,1}}&=&\big\{L_{-1}^4-10pL_{-2}L_{-1}^2+9p^2L_{-2}^2+2p(12p-5)L_{-3}L_{-1}
   -6p(6p^2-4p+1)L_{-4}\big\}\ket{\D_{4,1}}\nn
 \ket{\la_{5,1}}&=&\big\{L_{-1}^5-20pL_{-2}L_{-1}^3+64p^2L_{-2}^2L_{-1}
   +6p(14p-5)L_{-3}L_{-1}^2-64p^2(3p-1)L_{-3}L_{-2}\nn
  &&\quad-\ \!12p(24p^2-14p+3)L_{-4}L_{-1}+8p(3p-1)(24p^2-14p+3)L_{-5}\big\}\ket{\D_{5,1}}
\label{laD}
\eea
The singular vectors $\ket{\la_{1,s}}$ follow from these by application of the general relation
\be
 \ket{\la_{r,s}}=\ket{\la_{s,r}}\big|_{p\to 1/p}
\label{singsym}
\ee

\subsection{Nahm level, basis and spurious subspace}

The enveloping algebra of the Virasoro algebra contains the subalgebra 
\be
 \Ac_n=\big\langle\prod_{j=1}^m L_{-\ell_j};\ \sum_{j=1}^m\ell_j\geq n,\, m\in\mathbb{N}\big\rangle
\ee
generated by all products of Virasoro modes whose $L_0$ grading is greater than or equal to $n$
where $n\geq0$.
This subalgebra can be used to define a filtration of the Virasoro module $\Hc$ as the family
\be
 \Hc^n=\Hc/\Ac_{n+1}\Hc
\ee
of quotient spaces. Here we refer to the integer $n$ as the Nahm level.
At Nahm level $n\in\mathbb{N}_0$, the co-multiplications $\D(L_{-m})$ and 
$\Dti(L_{-m})$ vanish for $m>n$. At level $0$, all co-multiplications of negative
Virasoro modes thus vanish. It follows, in particular, that 
\be
 L_{-1}^i\times L_{-1}^j=(-1)^jL_{-1}^{i+j}\times I
\ee
at level $0$.
{}Furthermore, from 
\be
 \D(L_0)\big\{L_{-1}^\ell\times I\big\}\ket{\D}\times\ket{\D'}
 =\big\{L_{-1}^{\ell+1}\times I+\big(\D+\D'+\ell\big)L_{-1}^\ell\times I\big\}\ket{\D}\times\ket{\D'}
\label{L0l}
\ee
it follows by induction that
\be
 0=\big\{L_{-1}^k\times I\big\}\ket{\D}\times\ket{\D'}\quad\Rightarrow\quad
 0=\big\{L_{-1}^\ell\times I\big\}\ket{\D}\times\ket{\D'},\quad \ell\geq k
\label{lk0}
\ee

Let us consider the fusion product $(r,1)\otimes (1,s)$ assuming that $(1,s)$ is the
highest-weight module $Q_{1,s}$ while recalling that $(r,1)=Q_{r,1}$. 
As an {\em initial} vector space for the examination of this fusion product  at Nahm level $0$, 
we may consider
\be
 \Big\{\big\{L_{-1}^\ell\times I\big\}\ket{\D_{r,1}}\times\ket{\D_{1,s}};\ \ell\in\mathbb{Z}_{0,r-1}\Big\}
\label{basis}
\ee
or the similar set based on $s$ vectors of the form 
$\{I\times L_{-1}^\ell\}\ket{\D_{r,1}}\times\ket{\D_{1,s}}$.
Indeed, we can use the vanishing of the singular vector $\ket{\la_{r,1}}$ in
\be
 \ket{\la_{r,1}}\times\ket{\D_{1,s}}
\ee
to express the vector $\{L_{-1}^r\times I\}\ket{\D_{r,1}}\times\ket{\D_{1,s}}$ first in terms of
vectors of the form $\{(L_{-n_1}\ldots L_{-n_j})\times I\}\ket{\D_{r,1}}\times\ket{\D_{1,s}}$ where 
$j<r$ and subsequently in terms of the initial vectors (\ref{basis})
\be
 \big\{L_{-1}^r\times I\big\}\ket{\D_{r,1}}\times\ket{\D_{1,s}}
  =\Big\{\sum_{\ell=0}^{r-1}\al^r_\ell L_{-1}^\ell\times I\Big\}\ket{\D_{r,1}}\times\ket{\D_{1,s}}
\label{al}
\ee
The relation (\ref{L0l}) then
allows us to ignore $\{L_{-1}^k\times I\}\ket{\D_{r,1}}\times\ket{\D_{1,s}}$ for $k>r$.
For $r=2$, in particular, the decomposition (\ref{al}) reads
\be
 L_{-1}^2\times I=-pL_{-1}\times I+p\D_{1,s}I\times I
\ee 

The initial vector space $Q_{r,1}^s\otimes Q_{1,s}^n$
at Nahm level $n$ is constructed as the tensor product of the ``special subspace"
$Q_{r,1}^s=\{L_{-1}^\ell\ket{\D_{r,1}};\,\ell\in\mathbb{Z}_{0,r-1}\}$ of $Q_{r,1}$
and the space $Q_{1,s}^n=\{L_{-k_1}\ldots L_{-k_m}\ket{\D_{1,s}};\, k_1+\ldots+k_m\leq n\}$
of states up to Virasoro level $n$ in $Q_{1,s}$.
Depending on the model (labelled by $p$) and the second fusion factor $(1,s)$,
the linear span of the initial vector space may contain a spurious subspace \cite{GK9604} 
generated by linear relations on the initial vector space. 
The fusion space $(Q_{r,1}\otimes Q_{1,s})_{\mathrm{f}}^n$
at Nahm level $n$ is the complement to the spurious subspace of $Q_{r,1}^s\otimes Q_{1,s}^n$.

\subsection{Kac representation $(2,3)$ in critical dense polymers ${\cal LM}(1,2)$}
\label{App23}

We consider the fusion $(2,1)\otimes(1,3)=(2,3)$ for $p=2$.
The Kac representation $(2,1)$ is the irreducible highest-weight module 
$Q_{2,1}=M_{2,1}=\Vc(1)$ generated from the highest-weight vector $\ket{\D_{2,1}}$ where 
\be
 \ket{\la_{2,1}}=
 \big(L_{-1}^2-2L_{-2}\big)\ket{\D_{2,1}}=0,\qquad\D_{2,1}=1
\ee 
The Kac representation $(1,3)$ is the reducible yet indecomposable highest-weight module 
$Q_{1,3}$ generated from the highest-weight vector $\ket{\D_{1,3}}$ where 
\be
 \ket{\la_{1,3}}=
 \big(L_{-1}^3-2L_{-2}L_{-1}\big)\ket{\D_{1,3}}=0,\qquad\D_{1,3}=0
\label{sing13}
\ee 

Here we analyze the fusion $(2,1)\otimes(1,3)=(2,3)$ up to Nahm level 2 and find that
the eight-dimensional space $Q_{2,1}^s\otimes Q_{1,3}^2$ containing the fusion space 
$\big(Q_{2,1}\otimes Q_{1,3}\big)_{\mathrm{f}}^2$ of our interest also
contains a two-dimensional spurious subspace defined by the relations
\bea
 0&=&\big\{L_{-1}\times L_{-1}+ I\times L_{-1}^2\big\}\ket{\D_{2,1}}\otimes\ket{\D_{1,3}}\nn
 0&=&\big\{2I\times L_{-1}+2L_{-1}\times L_{-1}+4I\times L_{-2}-L_{-1}\times L_{-1}^2
   +2L_{-1}\times L_{-2}\big\}\ket{\D_{2,1}}\otimes\ket{\D_{1,3}}
\eea
Likewise, we find
\bea
 \big(Q_{2,1}\otimes Q_{1,3}\big)_{\mathrm{f}}^1&=&\big(Q_{2,1}^s\otimes Q_{1,3}^1\big)\big/
  \big(0=2\ket{\D_{2,1}}\otimes L_{-1}\ket{\D_{1,3}}+L_{-1}\ket{\D_{2,1}}\otimes 
   L_{-1}\ket{\D_{1,3}}\big)  \nn
 \big(Q_{2,1}\otimes Q_{1,3}\big)_{\mathrm{f}}^0&=&Q_{2,1}^s\otimes Q_{1,3}^0
\eea
The Virasoro generator $\D(L_0)$ is diagonalizable on the spaces
$\big(Q_{2,1}\otimes Q_{1,3}\big)_{\mathrm{f}}^n$, $n=0,1,2$, with suitably normalized 
eigenvectors given by
\bea
 \ket{0}^0&=& 3L_{-1}\ket{\D_{2,1}}\otimes \ket{\D_{1,3}}    \nn
 \ket{1}^0&=& \big\{I\times I+L_{-1}\times I\big\}\ket{\D_{2,1}}\otimes\ket{\D_{1,3}}   
\eea
\bea
 \ket{0}^1&=&-3\ket{\D_{2,1}}\otimes L_{-1}\ket{\D_{1,3}}    \nn
 \ket{1}^1&=& \big\{I\times I-L_{-1}\times I-2I\times L_{-1}\big\}
              \ket{\D_{2,1}}\otimes\ket{\D_{1,3}}   \nn
 \ket{2}^1&=&\big\{L_{-1}\times I+I\times L_{-1}\big\}
              \ket{\D_{2,1}}\otimes\ket{\D_{1,3}}    
\eea
and
\bea
 \ket{0}^2&=&\big\{I\times L_{-1}+L_{-1}\times L_{-1}+2I\times L_{-2}+L_{-1}\times L_{-2}\big\}
              \ket{\D_{2,1}}\otimes\ket{\D_{1,3}} \nn
 \ket{1}^2&=&\big\{2I\times L_{-1}+L_{-1}\times L_{-1}+3I\times L_{-2}+L_{-1}\times L_{-2}\big\}
              \ket{\D_{2,1}}\otimes\ket{\D_{1,3}} \nn
 \ket{2}^2_1&=&\big\{2I\times L_{-1}+2I\times L_{-2}+L_{-1}\times L_{-2}\big\}
              \ket{\D_{2,1}}\otimes\ket{\D_{1,3}} \nn
 \ket{2}^2_2&=&\big\{2I\times I-L_{-1}\times I-I\times L_{-1}-L_{-1}\times L_{-1}-2I\times L_{-2}\big\}
              \ket{\D_{2,1}}\otimes\ket{\D_{1,3}} \nn
 \ket{3}^2_1&=&\big\{-I\times I+L_{-1}\times I+2I\times L_{-1}+3I\times L_{-2}
  +L_{-1}\times L_{-2}\big\}
              \ket{\D_{2,1}}\otimes\ket{\D_{1,3}} \nn
 \ket{3}^2_2&=&\big\{-2I\times I+2L_{-1}\times I+2I\times L_{-1}
  +L_{-1}\times L_{-1}+2I\times L_{-2}\big\}
              \ket{\D_{2,1}}\otimes\ket{\D_{1,3}} 
\eea
When restricting from level 2 to level 1, it is $\ket{2}^2_2$ which becomes $\ket{2}^1$, while the
somewhat unusual normalizations of $\ket{0}^0$ and $\ket{0}^1$ follow from the normalization
of $\ket{0}^2$ under similar level restrictions.
The actions of the Virasoro generators $L_{\pm1}$ and $L_{\pm2}$ on these states are also
examined using the co-multiplication and we find that the non-trivial
actions read
\be
 \begin{array}{c}
 L_{-2}\ket{0}=3\ket{2}_1,\hspace{1cm} L_2\ket{2}_1=-\frac{1}{3}\ket{0},\hspace{1cm}
   L_1\ket{1}=-\frac{1}{3}\ket{0},\hspace{1cm} L_2\ket{2}_2=-\ket{0}
 \\[.3cm]
 L_{-1}\ket{1}=\ket{2}_2,\hspace{1cm}L_{-1}\ket{2}_2=\ket{3}_2,\hspace{1cm}
   L_{-2}\ket{1}=\ket{3}_1
 \\[.3cm]
 L_1\ket{2}_2=2\ket{1},\hspace{1cm}L_1\ket{3}_1=3\ket{2}_2,\hspace{1cm}
   L_1\ket{3}_2=6\ket{2}_2,\hspace{1cm}L_2\ket{3}_1=3\ket{1},\hspace{1cm}
   L_2\ket{3}_2=6\ket{1}
 \end{array}
\ee
where the level indications have been omitted.
With 
\bea
 \{\ket{0},\ket{2}_1\}&\subset&\Vc(0)\nn 
 \{\ket{1},\ket{2}_2,\ket{3}_1,\ket{3}_2\}\big/\big(0=\ket{3}_2-2\ket{3}_1\big)&\subset&\Vc(1)\nn
 \{\ket{3}_2-2\ket{3}_1\}&\subset&\Vc(3)
\eea
this is in accordance with the conjectured structure (\ref{23013}) 
of the non-highest-weight module $(2,3)$ appearing in the fusion $(2,1)\otimes(1,3)=(2,3)$.
It is noted that 
\be
 \ket{3}_{\mathrm{sing}}=\ket{3}_2-2\ket{3}_1
\ee
corresponds to the singular vector at Virasoro level 2 in $(1,5)$ 
from which the submodule $\Vc(3)$ is generated, and that
the reducible yet indecomposable highest-weight module $(1,5)$ admits the short exact sequence
\be
 0\to\Vc(3)\to(1,5)\to\Vc(1)\to0
\ee

\subsection{Fusion product $(1,3)\otimes(1,3)$ in critical dense polymers ${\cal LM}(1,2)$}
\label{App1313}

Here we analyze the fusion product $(1,3)\otimes(1,3)$ up to Nahm level 2 in ${\cal LM}(1,2)$.
We use the singular vector (\ref{sing13})
and find that the twelve-dimensional space $Q_{1,3}^s\otimes Q_{1,3}^2$ containing the 
fusion space $\big(Q_{1,3}\otimes Q_{1,3}\big)_{\mathrm{f}}^2$ also
contains a two-dimensional spurious subspace defined by the relations
\bea
 0&=&\big\{2L_{-1}\times L_{-1}-2L_{-1}\times L_{-1}^2-L_{-1}^2\times L_{-1}^2
   +L_{-1}\times L_{-2}+2L_{-1}^2\times L_{-2}\big\}\ket{\D_{1,3}}\otimes\ket{\D_{1,3}}\nn
 0&=&\big\{L_{-1}^2\times L_{-1}+ L_{-1}\times L_{-1}^2\big\}\ket{\D_{1,3}}\otimes\ket{\D_{1,3}}
\eea
Likewise, we find
\bea
 \big(Q_{1,3}\otimes Q_{1,3}\big)_{\mathrm{f}}^1&=&\big(Q_{1,3}^s\otimes Q_{1,3}^1\big)\big/
  \big(0=2L_{-1}\ket{\D_{1,3}}\otimes L_{-1}\ket{\D_{1,3}}
   +L_{-1}^2\ket{\D_{1,3}}\otimes L_{-1}\ket{\D_{1,3}}\big)  \nn
 \big(Q_{1,3}\otimes Q_{1,3}\big)_{\mathrm{f}}^0&=&Q_{1,3}^s\otimes Q_{1,3}^0
\eea
The conjectured decomposition of the fusion product $(1,3)\otimes(1,3)$ is described by the
structure diagram
\psset{unit=.25cm}
\setlength{\unitlength}{.25cm}
\be
\mbox{}
\hspace{-1cm}
 \mbox{
 \begin{picture}(20,6)(0,3.5)
    \unitlength=1cm
  \thinlines
\put(-3.5,1.5){$\R_1^1\oplus(1,5):$}
\put(1,2){$\Vc(1)$}
\put(-0.15,1){$\Vc(0)_0$}
\put(2,1){$\Vc(0)_1$}
\put(1.05,1){$\longleftarrow$}
\put(1.65,1.5){$\nwarrow$}
\put(0.65,1.5){$\swarrow$}
\put(4,1.5){$\bigoplus$}
\put(5.5,2){$\tilde{\Vc}(3)$}
\put(6.5,1){$\tilde{\Vc}(1)$}
\put(6.15,1.5){$\nwarrow$}
 \end{picture}
}
\ee
The indices on the irreducible highest-weight modules $\Vc(0)$ and the tildes on 
the irreducible subfactors of $(1,5)$ are immaterial
but introduced for ease of reference to the different modules.
The rank-2 module $\R_1^1$ is generated by the action of the Virasoro modes
on the vectors $\ket{0}_0$ and $\ket{0}_1$ where
\be
 L_0\ket{0}_1=\ket{0}_0
\ee
while the highest-weight module $(1,5)$ is generated from the vector $\tilde{\ket{1}}$.
At Nahm level 2, we should then recover
\bea
 \{\ket{0}_0,L_{-2}\ket{0}_0\}&\subset&\Vc(0)_0\nn 
 \{\ket{0}_1,L_{-2}\ket{0}_1\}&\subset&\Vc(0)_1\nn 
 \{L_{-1}\ket{0}_1,L_{-1}^2\ket{0}_1\}&\subset&\Vc(1)
\eea
and
\bea
 \{\tilde{\ket{1}},L_{-1}\tilde{\ket{1}},L_{-1}^2\tilde{\ket{1}},L_{-2}\tilde{\ket{1}}\}
   \big/\big(0=(L_{-1}^2-2L_{-2})\tilde{\ket{1}}\big)&\subset&\tilde{\Vc}(1)\nn
 \{(L_{-1}^2-2L_{-2})\tilde{\ket{1}}\}&\subset&\tilde{\Vc}(3)
\eea
satisfying
\be
 L_0L_{-2}\ket{0}_1=2L_{-2}\ket{0}_1+L_{-2}\ket{0}_0,\qquad
 L_2(L_{-1}^2-2L_{-2})\tilde{\ket{1}}=0
\ee
in particular.
This is confirmed when considering
\bea
 \ket{0}_0&=&\big(0,0,0,0,1,0,-1,0,2,1\big)\nn
 L_{-2}\ket{0}_0&=&\big(0,0,0,0,6,0,0,0,6,3\big)\nn
  \ket{0}_1&=&\big(-3,2,-\tfrac{1}{2},0,1,0,0,0,1,0\big)+\al\ket{0}_0\nn
 L_{-2}\ket{0}_1&=&\big(0,-6,3,0,-5,0,0,-3,-2,-\tfrac{5}{2}\big)+\al L_{-2}\ket{0}_0\nn
 L_{-1}\ket{0}_1&=&\big(0,0,0,-3,0,0,\tfrac{3}{2},0,-3,0\big)\nn
 L_{-1}^2\ket{0}_1&=&\big(0,-6,3,0,0,-3,-3,0,6,0\big)
\label{00}
\eea
and
\bea
 \tilde{\ket{1}}&=&\big(0,0,0,-2,8,0,-3,0,10,4\big)\nn
 L_{-1}\tilde{\ket{1}}&=&\big(0,4,-2,0,-4,-2,2,0,-4,0\big)\nn
 L_{-1}^2\tilde{\ket{1}}&=&\big(0,-8,8,0,8,0,-4,0,8,0\big)\nn
 L_{-2}\tilde{\ket{1}}&=&\big(0,-4,4,0,8,0,0,0,12,4\big)
\eea
in the ordered basis
\be
\begin{array}{c}
  \big\{ I\times I,\,L_{-1}\times I,\,L_{-1}^2\times I,\,I\times L_{-1},\,L_{-1}\times L_{-1},\,
  I\times L_{-1}^2,\,L_{-1}\times L_{-1}^2,
\\[.2cm]
   I\times L_{-2},\, L_{-1}\times L_{-2},\,L_{-1}^2\times L_{-2}\big\}
 \ket{\D_{1,3}}\otimes\ket{\D_{1,3}}
\end{array}
\ee
The parameter $\al$ in (\ref{00}) is free and corresponds to a gauge transformation.

\section{Kac fusion algebra for critical dense polymers ${\cal LM}(1,2)$}
\label{AppCrit}

The Kac fusion algebra for critical dense polymers ${\cal LM}(1,2)$ satisfies
\be
 \big\langle (r,s);\, r,s\in\mathbb{N}\big\rangle
 =\big\langle(r,2),\, (r,2j-1),\,\R_r;\,r,j\in\mathbb{N}\big\rangle
\ee
where $\R_r=\R_r^1$. The fusion rules are
\bea
 (r,2)\otimes(r',2)
 \!\!&=&\!\!\bigoplus_{\ell=|r-r'|+1,\,\mathrm{by}\,2}^{r+r'-1}\R_\ell   
  \nn
 (r,2)\otimes(r',2j'-1)
 \!\!&=&\!\!\bigoplus_{\ell=|r-r'|+1,\,\mathrm{by}\,2}^{r+r'-1}
  \   \bigoplus_{k=|\ell-j'+\frac{1}{2}|+\frac{1}{2}}^{\ell+j'-1} (k,2)
  \nn
 (r,2)\otimes\R_{r'}
 \!\!&=&\!\!\bigoplus_{\ell=|r-r'|}^{r+r'}\big(2-\delta_{\ell,|r-r'|}-\delta_{\ell,r+r'}\big)(\ell,2)
  \nn
 (r,2j-1)\otimes(r',2j'-1)
 \!\!&=&\!\!\bigoplus_{\ell=|r-r'|+1,\,\mathrm{by}\,2}^{r+r'-1}\
     \bigoplus_{k=|j-j'|+1,\,\mathrm{by}\,2}^{j+j'-3}\
     \bigoplus_{i=|\ell-k|+1,\,\mathrm{by}\,2}^{\ell+k-1}\!\!\R_i
   \oplus\bigoplus_{\ell=|r-r'|+1,\,\mathrm{by}\,2}^{r+r'-1}\!\!(\ell,2j+2j'-3)
  \nn
 (r,2j-1)\otimes\R_{r'}
 \!\!&=&\!\!\bigoplus_{\ell=|r-r'|+1,\,\mathrm{by}\,2}^{r+r'-1}
  \ \bigoplus_{k=|\ell-j+\frac{1}{2}|+\frac{1}{2}}^{\ell+j-1}\R_k
  \nn
 \R_r\otimes\R_{r'}
 \!\!&=&\!\!\bigoplus_{\ell=|r-r'|}^{r+r'}\big(2-\delta_{\ell,|r-r'|}-\delta_{\ell,r+r'}\big)\R_\ell
\eea
where it is noted that some summations are in steps of 1 while others are in steps
of 2. The vertical fusion rule
\be
 (1,2j-1)\otimes(1,2j'-1)=\bigoplus_{\ell=|j-j'|+1,\,\mathrm{by}\,2}^{j+j'-3}\R_\ell
  \oplus(1,2j+2j'-3)
\ee
revises the similar formula in \cite{PR0610}.

\section{Jordan canonical form of the hamiltonian $H_{s,s'}^{(N)}$}
\label{AppJordan}

Let us introduce the $d\times d$ square matrix $D=D^{(d)}$
\be
D=\left(\!\!\begin{array}{ccccccc}
0&1&&&&&\\
1&0&1&&&&\\
&1&0&1&&&\\
&&1&0&&&\\
&&&& \ddots&&\\
&&&& &0&1\\
&&&& &1&0
\end{array}\!\!\right)
\label{D}
\ee
It has $d$ eigenvalues
\be
 \al_j=2\cos\frac{j\pi}{d+1},\ \ \ \ \ \ \ j\in\mathbb{Z}_{1,d}
\ee
all of which are distinct. The associated eigenvectors can be normalized as
\be
 v_j=\left(\!\begin{array}{c}
 \sin\frac{j\pi}{d+1}\\ \vdots \\  \sin\frac{ij\pi}{d+1}\\ \vdots \\ \sin\frac{dj\pi}{d+1}\end{array}\!\right),
  \ \ \ \ \ \ \ i,j\in\mathbb{Z}_{1,d}
\label{vj}
\ee
The matrix $D$ is diagonalized by $K=K^{(d)}$ constructed by concatenating the
eigenvectors (\ref{vj}). That is,
\be
 K^{-1}DK=\mathrm{diag}(\al_1,\ldots,\al_d)
\ee
where
\be
 K_{ij}=\sqrt{\frac{2}{d+1}}\sin\frac{ij\pi}{d+1},\ \ \ \ \ \ \ i,j\in\mathbb{Z}_{1,d}
\ee
satisfying
\be
 K^{-1}=K^t=K
\ee

We can construct the hamiltonian $H_{s,s'}^{(N)}$ discussed in
Section~\ref{SecEviFusionLattice} as the block matrix
\be 
 -H_{s,s'}^{(N)}=\begin{pmatrix} D_\beta&\delta^{(N-1)}\\ 0_{1\times (N-1)}&0\end{pmatrix}
\ee
where $D_\beta=D^{(N-1)}_\beta$ is the $(N-1)\times(N-1)$ square matrix 
\be
 D_\beta=D+\beta I,\qquad\beta=-2\cos\tfrac{\pi}{p}
\ee
while the entries of the $(N-1)$-vector $\delta^{(N-1)}$ are
\be
 \delta^{(N-1)}_j=\delta_{j,s-t-1},\qquad j\in\mathbb{Z}_{1,N-1}
\ee
The matrix $-H_{s,s'}^{(N)}$ is similar to
\be
 -\bar{H}_{s,s'}^{(N)}=-\bar{K}^{-1}H_{s,s'}^{(N)}\bar{K},\qquad
 \bar{K}=\begin{pmatrix} K&0_{(N-1)\times1}\\0_{1\times(N-1)}&1\end{pmatrix}
\ee
that is,
\be
 -\bar{H}_{s,s'}^{(N)}=\left(\!\!\begin{array}{cccccc}
  \beta+2\cos\frac{\pi}{N}&0&\dots&\dots&0&\sqrt{\frac{2}{N}}\sin\frac{(s-t-1)\pi}{N}\\
  0&\ddots&&&\vdots&\vdots\\
  \vdots&&\beta+2\cos\frac{j\pi}{N}&&\vdots&\sqrt{\frac{2}{N}}\sin\frac{j(s-t-1)\pi}{N}\\
  \vdots&&&\ddots&0&\vdots\\
  \vdots&&&&\beta+2\cos\frac{(N-1)\pi}{N}&\sqrt{\frac{2}{N}}\sin\frac{(N-1)(s-t-1)\pi}{N}\\
  0&\dots&\dots&\dots&0&0   \end{array}\!\!\right)
\ee
Since two similar matrices have the same Jordan canonical form,
this simple result facilitates a straightforward analysis of the Jordan decomposition
of $H_{s,s'}^{(N)}$ itself.
{}From $\beta+2\cos\tfrac{j\pi}{N}\neq\beta+2\cos\tfrac{j'\pi}{N}$ for $j\neq j'$, it follows
that $H_{s,s'}^{(N)}$ is diagonalizable if $\beta\neq-2\cos\tfrac{j\pi}{N}$ for all
$j\in\mathbb{Z}_{1,N-1}$. It is also readily seen that $H_{s,s'}^{(N)}$ is diagonalizable
if $\beta+2\cos\tfrac{j_0\pi}{N}=\sin\tfrac{j_0(s-t-1)\pi}{N}=0$ for some
$j_0\in\mathbb{Z}_{1,N-1}$, while $H_{s,s'}^{(N)}$ is non-diagonalizable if there 
exists $j_0\in\mathbb{Z}_{1,N-1}$ for which
$\beta=-2\cos\tfrac{j_0\pi}{N}$ and $\sin\tfrac{j_0(s-t-1)\pi}{N}\neq0$.


\end{document}